\documentclass[a4paper,fleqn,usenatbib]{mnras}
\usepackage{newtxtext,newtxmath}
\usepackage{graphicx}

\title[AGB Variables in Sgr dIG]{A Remarkable Oxygen-rich Asymptotic Giant Branch Variable in the Sagittarius Dwarf Irregular Galaxy}

\author[Whitelock et al.]{Patricia A. Whitelock$^{1,2}$, 
John W. Menzies$^1$,  Michael W. Feast$^{2,1}$   and Paola Marigo$^3$\\
      $^1$ South African Astronomical Observatory, P.O.Box 9, 7935
           Observatory, South Africa.\\
      $^2$ Department of Astronomy,
           University of Cape Town, 7701 Rondebosch, South Africa.\\
       $^3$ Department of Physics and Astronomy G. Galilei, University of Padova, Vicolo dellÕOsservatorio 3, I-35122 Padova, Italy. }

\begin{document}
\maketitle
\begin{abstract} We report and discuss $JHK_S$ photometry for Sgr~dIG, a very metal-deficient galaxy in the Local Group, obtained over 3.5 years with the Infrared Survey Facility in South Africa. Three large amplitude asymptotic giant branch variables are identified. One is an oxygen-rich star that has a pulsation period of 950 days, that was until recently undergoing hot bottom burning, with $M_{bol}\sim-6.7$. It is surprising to find a variable of this sort in Sgr~dIG, given their rarity in other dwarf irregulars. Despite its long period the star is relatively blue and is fainter, at all wavelengths shorter than $4.5 \mu$m, than anticipated from period-luminosity relations  that describe hot bottom burning stars. A comparison with models suggests it had a main sequence mass $M_i \sim 5\, M_{\odot}$ and that it is now near the end of its AGB evolution. The other two periodic variables are carbon stars with periods of 670 and 503 days ($M_{bol}\sim-5.7$ and $-5.3$). They are very similar to other such stars found on the AGB of metal deficient Local Group Galaxies and a comparison with models suggests $M_i \sim 3\, M_{\odot}$. We compare the number of AGB variables in Sgr~dIG to those in NGC\,6822 and IC\,1613, and suggest that the differences may be due to the high specific star formation rate and low metallicity of Sgr~dIG.

\end{abstract}
\section{Introduction}\label{intro}

The Sagittarius Dwarf Irregular Galaxy (Sgr~dIG) is one of the outermost members of the Local Group. It is a particularly difficult galaxy to observe as the field ($l=21^{\rm o}$,  $b=-16^{\rm o}$) is dominated by foreground stars from the Galactic disk and bulge. Nevertheless, it has come under increasing scrutiny in recent years, primarily because of its extremely low metallicity (the most recent measurement  being $ \rm [Fe/H]=-1.88^{+0.13}_{-0.09}$ \citep{Kirby2017}, although earlier estimates were lower), which  makes it more representative of the early universe than other systems in which individual stars can be studied. The oxygen abundance measured from H{\sc ii} regions is also low, with $\rm 12+\log (O/H)$ in the range 7.26 to 7.50 \citep{Saviane2002}. 
\\
The young blue stars are concentrated towards the centre of Sgr~dIG while the redder stellar populations are more widely distributed and the H{\sc i} gas covers a much bigger volume \citep[][and references therein]{Beccari2014, Higgs2016}. The star formation history \citep{Weisz2014} suggests an extended period of star formation and is similar to that of other dwarf irregulars, although it has the highest gas fraction of any galaxy in the Local Group and a particularly high specific star formation rate, making it one of the fastest growing galaxies in the Local Group according to \citet{Kirby2017}. While it appears to be isolated it shows signs of interaction \citep{Higgs2016},  including dusty AGB candidates in its outer regions  \citep{Cook1988, McQuinn2017}. 
\\
 \citet{Demers2002} identified C stars via narrow band optical photometry and \citet{Gullieuszik2007} via $JHK_S$ photometry. Most of these will be AGB stars. \citet{Boyer2015a, Boyer2015b} identified a number of upper AGB candidates on the basis of their positions in a Spitzer (3.6 and 4.5 $\mu$m) colour-magnitude diagram and/or their variability. 

For the Sgr~dIG \citet{Momany2005} give $E(B-V)=0.12 \pm 0.05$ and find $(m-M)_0=25.10\pm0.11$, \citet{Beccari2014} use $E(B-V)=0.107 \pm 0.10$ and get $(m-M)_0=25.16\pm0.11$, while \citet{Higgs2016}  and \citet{McQuinn2017} find $(m-M)_0=25.36\pm0.15$ and $25.18\pm0.04$, respectively; all estimates are from the red giant branch tip (RGBT).
In the following we assume a distance modulus of 25.2 and an interstellar extinction of $A_V=0.34$, which results in $A_J=0.1 $, $A_H=0.06$ and $A_K=0.04$, but our findings are not sensitive to the interstellar extinction. Where we make use of relations defined for the LMC we assume its distance modulus to be 18.5.

This work forms part of a broad study of AGB variables in Local Group galaxies which so far has covered the dwarf spheroidals: Leo I \citep{Menzies2002, Menzies2010}, Phoenix \citep{Menzies2008}, Fornax \citep{Whitelock2009} and Sculptor \citep{Menzies2011}, as well as two dwarf irregulars, NGC 6822 \citep{Whitelock2013} and IC\,1613 \citep{Menzies2015}. 

Stars up to about $10\, M_{\odot}$ are thought to undergo AGB evolution, where they reach very high luminosities while undergoing large amplitude pulsations. Dredge-up on the thermally pulsing AGB will increase the abundance of atmospheric carbon, turning normal O-rich stars into C-rich stars once C/O exceeds unity. The lower the initial O-abundance the more rapid the transition to C-rich star will be. However, it is now clear that AGB stars above a certain mass will undergo hot bottom burning (HBB), and the dredged-up carbon is burned to nitrogen, so the stars become O-rich again. The HBB process produces additional luminosity so these stars will be brighter than the core-mass luminosity relation predicts. Unfortunately, even the broad details of nucleosynthesis, dredge up and HBB, as well their dependence on initial metallicity, remain very uncertain \citep{Doherty2017, Karakas2017}.  Even the upper limit to the mass range for AGB evolution is uncertain, and is usually quoted at around 8 to 12 $M_{\odot}$ for the most massive super-AGB stars. Understanding these stars is vital for establishing the mass boundary between stars that will produce supernovae and those that end as white dwarfs and its dependence on abundance.
 It is therefore of particular interest to isolate luminous large-amplitude variables in a variety of environments with the objective of identifying HBB and/or super-AGB stars that can be studied in more detail. 


\section{Mira Period Luminosity Relation (PLR)}\label{Mira_PLR}

In addition to finding luminous large amplitude variables, one of the main objectives of our studies is calibrating and testing the PLR for Mira variables in different environments, in particular for Miras because these are  easily identified as large amplitude luminous stars that  potentially rival Cepheids as distance indicators \citep[][and other works in our series on Local Group galaxies]{Feast2013, Whitelock2013b, Whitelock2014}. We expect increasing interest in these variables and the PLR following commissioning of the James Web Space Telescope and with the use of adaptive optics in the next generation of ground based extremely large telescopes. 
 
Early work on the PLR  \citep[e.g.,][]{Feast1989,Hughes1990} was based on observations of LMC Miras and showed a clear relation at near-infrared wavelengths,  as well as for the bolometric magnitude.  Separate relations were derived for O- and C-rich variables, although the differences between these were small. For distance scale purposes the $K$ PLR appears to be the best; the amplitude is less than at shorter wavelengths as are the effects of interstellar and circumstellar reddening. It was clear even at this early stage that longer period ($P>420$ days) O-rich Miras are brighter at $K$ than a linear PLR relation fitted to shorter period stars would predict. 

\citet{Wood1999} showed that the Mira sequence represents fundamental pulsation and that semi-regular AGB variables also obey PLRs, but many of them pulsate in harmonics instead of, or as well as, in the fundamental mode. This work has subsequently been consolidated and extended by several other groups \citep[e.g.,][]{
Ita2004, Soszynski2009}. 

 It is well known that the period ($P$) of a radially pulsating star is a function of its density ($\rho$) and therefore of its current mass through the relation $P \sqrt {\rho} = constant$. 
 Observations of Globular Clusters \citep{Feast2002} and the kinematics of Miras  \citep{Feast2000, Feast2006} indicate that $P$ is also a function of the star's initial mass, 
from which we can deduce that, at least among short period Miras, there can be little evolution of period once a star becomes a Mira (i.e. a fundamental mode pulsator). \citet{Wood2015}  discussed  the possible extent of evolution within the PLR. 
 The fundamental (Mira) PLR can be understood as the locus of the end points of AGB evolution of stars of different initial mass and the period is  a function of both the initial mass and the current mass. The PLR can also be understood as a consequence of the core-mass luminosity relation.
 
As more Miras were discovered via infrared surveys \citep[e.g.,][]{Cioni2001, Ita2004,  Menzies2010, Menzies2008, Menzies2011, Menzies2015, Whitelock2009, Whitelock2013} it became clear that many C-rich Miras fall below the $K$ PLR, even at relatively short period.  This is a consequence of high circumstellar reddening and almost all of these reddened stars do fall on the bolometric PLR as our cited work on Local Group galaxies shows.

 \citet{Whitelock2003} suggested that the bright long period O-rich stars are actually HBB and this is why they are more luminous than the core-mass luminosity relation would predict. They further suggested that very long period Miras, i.e., OH/IR stars with $P>1000$ days, lie on the bolometric PLR, as they are no longer HBB (the bolometric luminosities of these very dusty stars are difficult to establish with very limited mid-infrared data).   The first part of this conclusion, that the stars are HBB, seems sound. The second part, that the very long period stars fall on the extrapolated PLR, is not entirely consistent with new work that includes more detailed mid-infrared studies.   

\citet{Ita2011} discuss the LMC PLRs at various wavelengths from  the near- to the mid-infrared and conclude that there is a kink in the PLRs between 400 and 500 days, such that the relation is steeper in the longer wavelength range.  More recently \citet{Yuan2017} have discussed essentially the same LMC data (and similar data for M33) and fitted a quadratic PLR as a better representation of the slope change.    The differences between the \citeauthor{Yuan2017} quadratic and the two straight lines of \citeauthor{Ita2011} is not significant in most practical terms, given the uncertainty associated with measured magnitudes. The clear interpretation of the multiple linear PLRs, in terms of fundamental and harmonic oscillations \citep{Wood1999, Wood2015},  would make it surprising if the PLR were non-linear over the short period range. At longer periods it is not clear how HBB affects the structure of the star and hence the pulsation period as well as the luminosity. Therefore it is not yet possible to predict whether the PLR should be linear or not.

\citet{Riebel2015} derived PLRs for multiply observed (so that mean magnitudes can be discussed) Magellanic Cloud Miras at $[3.6]$ and $[4.5]$, but omitted the so-called extreme AGB stars (x-AGB). These x-AGB stars have thick dust shells and are variously defined as those with $J-[3.6]>3.1$, $J-K>2$ or $[3.6]-[4.5]>  0.15$ \citep{Riebel2015, Boyer2015a}.  \citet{Whitelock2017} used the \citeauthor{Riebel2015} data, but included the x-AGB stars in their PLR analysis. For C-rich Mira variables they found a more complex relation with a colour dependency. It is beyond the scope of this paper to discuss this in more detail, and further work is needed to understand and characterise the behaviour of the longer period Miras. It seems likely we may have to separate those with periods significantly above 400 days in ways that depend on other parameters. 
 
In the following we discuss the AGB variables in the same terms as we have in our earlier work on Local Group galaxies while using the \citeauthor{Ita2011} and \citeauthor{Yuan2017} relationships to deal with HBB stars.

\begin{figure}
\includegraphics[width=8.5cm]{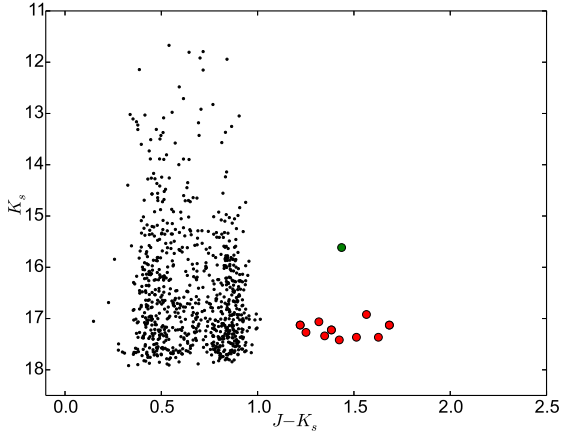}
\caption{Colour-magnitude diagram for Sgr~dIG. Red symbols show known or suspected C stars, the green symbol is the large amplitude variable, V1.   Note that V2 and V3 are not shown as we do not have $J$ for them, but both would be off scale to the right}
\label{fig_cm}
\end{figure}

\begin{figure}
\includegraphics[width=8.5cm]{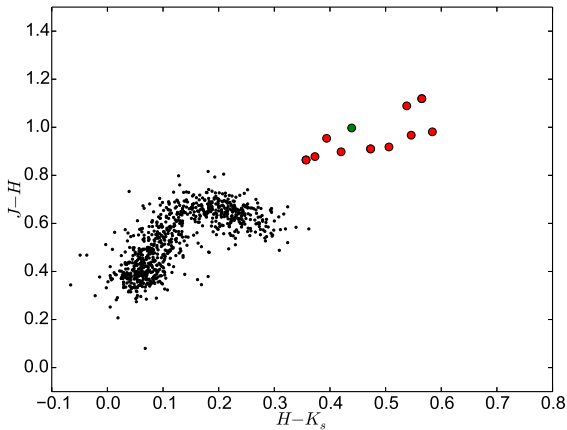}
\caption{Two colour diagram for Sgr~dIG. Coloured symbols as in Fig.~\ref{fig_cm}.  Note that V2 and V3 are not shown as we do not have $J-H$ for them, but both would be off scale; for V3 $H-K_S=1.56$. }
\label{fig_jhhk} 
\end{figure}

\section{Photometry}
Observations were made with the SIRIUS camera on the  InfraRed Survey Facility (IRSF) at Sutherland. The camera produces  simultaneous $J$, $H$ and $K_S$ images covering a $7.68 \times 7.68$ arcmin square field with a scale of 0.45 arcsec/pixel. The Sgr~dIG is sufficiently small to require only one pointing per image centred at $\alpha(2000.0)=$19:29:58.0, $\delta(2000.0)=$-17:40:41.0. The aim of the observational series was to find long-period variables; observations were made at 17 epochs spread over 3.5 years. For this field, 10 dithered images were combined after flat-fielding and dark and sky subtraction. Exposures were  of either 40 or 60 seconds' duration, depending on the seeing and on the brightness of the sky in the $K_S$ band. Photometry was performed using DoPHOT in `fixed-position' mode, using the best-seeing $H$-band image as a template. Small differences in pointing amongst the images meant that the area in common to them all for photometry is only $7.2 \times 7.2$ arcmin square. Aladin was used to correct the WCS on each template and RA and Dec were determined for each measured star. This allowed a cross-correlation to be made with the 2MASS catalogue \citep{Cutri2003}, and photometric zero points were determined by comparison of our photometry with that of 2MASS.


Fig.~\ref{fig_cm} shows the $K_S-(J-K_S)$ diagram and Fig.~\ref{fig_jhhk} the $(J-H)-(H-K_S)$
diagram for stars with standard deviations less than 0.2~mag in each band.  These diagrams have not been corrected for interstellar reddening which will amount to only $\Delta (J-H)=0.04$,  $\Delta (H-K_S)=0.02$ and  $\Delta (J-K_S)=0.06$. As anticipated most of the stars are foreground. The {\sc trilegal} \citep{Girardi2005} code (version 1.6) indicates that we would expect about 500 to 520 foreground stars from the Galaxy with $11<K_S<17.5$ (typically all with $J-K_S<1.1$), compared with the 730 actually observed in this area. This is fewer foreground stars than we might expect given that \citet{Gullieuszik2007} estimate contamination using a nearby field, away from Sgr~dIG, and conclude that the stars brighter than $K_S = 19.5$ and bluer than $J-K_S = 1.1$ are most probably foreground and that all the stars redder than this limit are candidate C stars belonging to Sgr~dIG.  

The RGBT will be at $K_S\sim 19.6$.  All stars with $J-K_S > 1.1$ are brighter than this and therefore probably on the AGB; they fall on the red plume in Fig.~\ref{fig_jhhk}. Possible contaminants are brown dwarfs (the {\sc trilegal} simulations produced a brown dwarf with colours and magnitudes that overlap with the C stars in one of fourteen simulations) and unresolved background galaxies (although their $J-H$ is usually slightly smaller than that of C stars with the same $H-K_S$ \citep{Whitelock2009}). The potential contamination of the red stars is very small and neither brown dwarfs nor background galaxies will be large amplitude variables.  

A comparison with IC\,1613, a closer dwarf irregular, (figs. 3 and 4 from \citet{Menzies2015}) shows that the brightest Sgr~dIG source is slightly redder  and brighter than the four HBB sources in IC\,1613 while the other Sgr~dIG sources fall in the region of small amplitude C-rich variables.
 
\begin{table}
\centering
\caption{$HK_S$ Photometry of the variables from ESO images}
\begin{tabular}{cccccccc}

JD & mag & err & mag & err & Mag & err\\
 & \multicolumn{2}{c}{V1}  & \multicolumn{2}{c}{V2}  & \multicolumn{2}{c}{V3}\\
\multicolumn{5}{l}{\bf $K_S$} \\
1037.120 & 15.60 & 0.05 & 17.46 & 0.06 & 15.70 & 0.05\\
2851.998 & 16.06 & 0.05 & 17.25 & 0.14 & 17.21 & 0.10\\
4002.087 & 16.37 & 0.05 & 18.28 & 0.10 & 16.17 & 0.05\\
4341.033 & 15.48 & 0.05 & 17.34 & 0.09 & 16.94 & 0.05\\
\multicolumn{5}{l}{\bf $H$} \\
1036.999 & 16.07 & 0.10 & -- & -- & -- & --\\
1038.052 & 16.12 & 0.10 & -- & -- & -- & --\\
3186.247 & 15.76 & 0.05 & -- & -- & -- & --\\
3554.278 & 16.24 & 0.12 & -- & -- & -- & --\\
5755.278 & 16.18 & 0.08 & -- & -- & -- & --\\

\end{tabular}\label{tab_extraV}
\end{table}

\subsection{Archival near-infrared photometry}
\begin{figure}
\center
\includegraphics[width=7cm]{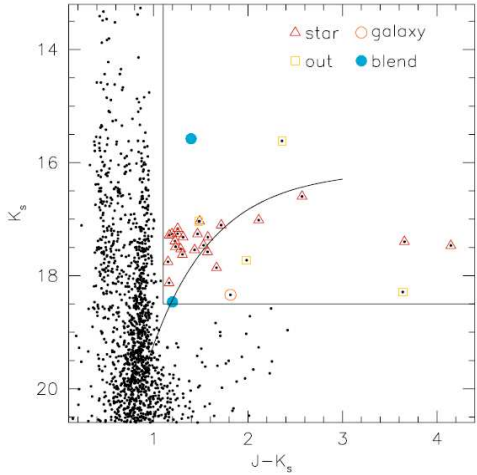}
\caption{Colour-magnitude diagram of \citet{Gullieuszik2007}. The curved line is the colour magnitude relation for carbon stars from \citet{Totten2000} and the box shows the area in which C stars were found according to the selection criteria of \citeauthor{Gullieuszik2007}. }
\label{fig_Gull}
\end{figure}

\citet{Gullieuszik2007} searched for carbon stars in Sgr~dIG using $V$, $J$ and $K_S$ photometry and their colour-magnitude diagram looks similar to ours, but goes considerably deeper\footnote{their $JK_S$ photometry is from the 3.58-m NTT, whereas ours is from the 1.4-m IRSF.}, as can be seen in Fig.~\ref{fig_Gull}. The stars redder than $J-K_S = 2.0$ are too faint for us to measure in all colours, though most are visible on our $K_S$ images. 

The bright star indicated as a blend in Fig.~\ref{fig_Gull} is our M type variable, V1, discussed in detail below.  \citet{Gullieuszik2007} claim that V1 was blended on their images, presumably on the basis of a large SHARP parameter in their DAOPHOT photometry, though they do not show the distribution of this parameter for Sgr~dIG. We therefore obtained the images used by \citet {Momany2005}, who did not give any photometry for the star, from the HST archive.  To extend the time span for calculation of the periods of the variables, we also obtained the SOFI $K_S$ and $H$ images from the ESO archive. Aperture photometry was performed on the variables  discussed below and the magnitudes were put on the same scale as our photometry. Incidentally, we obtained essentially the same result for $K_S$ as \citet{Gullieuszik2007} for V1 (number 21 in their table 4). The extra photometry is shown in Table~\ref{tab_extraV}.

\begin{table*}
\centering
\caption{IRSF $JHK_S$ Photometry of Large Amplitude Variables}
\begin{tabular}{cccccccccccccccc}
&\multicolumn{6}{c}{V1} & \multicolumn{2}{c}{V2} & \multicolumn{4}{c}{V3}\\
JD & $J$ & err($J$) & $H$ & err($H$)& $K_S$ & err($K$)& $K_S$ & err($K$) & $H$ & err($H$)& $K_S$ & err($K$)\\
-2450000 &\multicolumn{12}{c}{mag} \\
2352.497 & 16.87  & 0.02  & 15.86  & 0.01  & 15.45  & 0.02  &   --         & --   &  18.00 & 0.11&  16.55 & 0.06  \\
2440.578 & 16.88  & 0.02  & 15.92  & 0.01  & 15.45  & 0.01  &   17.28 & 0.15 &  17.51 & 0.06&  15.97 & 0.05  \\
2441.751 & 16.93  & 0.02  & 15.93  & 0.01  & 15.48  & 0.02  &   17.40 & 0.11 &  17.44 & 0.07&  16.10 & 0.05  \\
2507.259 & 17.01  & 0.02  & 16.08  & 0.01  & 15.63  & 0.01  &   16.98 & 0.11 &  17.41 & 0.07&  16.03 & 0.04  \\
2524.299 & 17.00  & 0.02  & 16.12  & 0.01  & 15.72  & 0.02  &   16.56 & 0.07 &  17.50 & 0.08&  15.96 & 0.04  \\
2775.579 & 17.53  & 0.03  & 16.45  & 0.02  & 15.95  & 0.02  &   16.59 & 0.10 &  18.46 & 0.12&  16.66 & 0.06  \\
2808.437 & 18.11  & 0.14  & 16.83  & 0.08  & 16.18  & 0.06  &   --    &--    &  --    & --  &  --    & --    \\
2809.348 & 17.63  & 0.04  & 16.47  & 0.02  & 16.07  & 0.03  &   17.39 & 0.11 &  18.23 & 0.14&  16.77 & 0.07  \\
2882.312 & 17.58  & 0.01  & 16.51  & 0.01  & 16.18  & 0.02  &   17.32 & 0.09 &  18.16 & 0.09&  16.70 & 0.07  \\
3173.407 & 16.72  & 0.01  & 15.78  & 0.01  & 15.34  & 0.01  &   16.65 & 0.06 &  18.38 & 0.12&  16.82 & 0.06  \\
3236.497 & 16.73  & 0.01  & 15.80  & 0.01  & 15.33  & 0.01  &   16.80 & 0.07 &  18.76 & 0.11&  16.95 & 0.06  \\
3259.218 & 16.81  & 0.01  & 15.86  & 0.01  & 15.39  & 0.01  &   --    & --   &  18.58 & 0.11&  16.92 & 0.06  \\
3260.220 & 16.77  & 0.01  & 15.86  & 0.01  & 15.41  & 0.01  &   16.66 & 0.07 &  18.77 & 0.13&  16.88 & 0.06  \\
3262.261 & 16.76  & 0.01  & 15.88  & 0.01  & 15.40  & 0.01  &   16.86 & 0.08 &  18.46 & 0.12&  16.91 & 0.07  \\
3293.239 & 16.88  & 0.01  & 15.94  & 0.01  & 15.49  & 0.01  &   16.35 & 0.08 &  18.33 & 0.11&  16.73 & 0.07  \\
3532.437 & 17.23  & 0.03  & 16.22  & 0.01  & 15.76  & 0.02  &   --   & --    &  17.53 & 0.07&  --    & --    \\
3616.307 & 17.52  & 0.01  & 16.44  & 0.01  & 15.98  & 0.01  &   17.43 & 0.13 &  18.04 & 0.07&  16.38 & 0.04  \\

\end{tabular}\label{tab_Var}
\end{table*}

\section{Variable Stars}\label{var_stars}
\begin{figure}
\includegraphics[width=8.5cm]{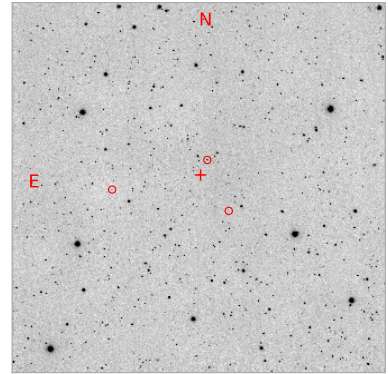}
\caption {Finding chart for the 3 variable stars using the template $K_S$ image. The variables are circled, and the centre of the galaxy is marked by a cross. The field shown is 7.68 arcmin square. }
\label{fig_Varchart}
\end{figure}

We are limited by sensitivity, and only one large amplitude variable was  found amongst our photometric sample. However, we performed aperture photometry on our $K_S$ frames at the positions of the reddest stars from the \citet{Gullieuszik2007} sample, their numbers (GRCHH) 2, 9, 14, 19, 24 and 29.  Both GRCHH\,14 and GRCHH\,19 have large amplitudes but are not obviously periodic, GRCHH\,9 may be a small amplitude irregular variable, while GCRHH\,29 is not visible on any of our $K_S$ frames. GRCHH\,2 and 24 appear regular with large amplitudes. The three large amplitude variables are listed in Table~\ref{tab_var2}, together with mid-infrared photometry from Spitzer \citep{Boyer2015a,Boyer2015b}. The spectral type for the M star comes from Boyer et al. (2017 in preparation) and is based on HST near-infrared colours. \citet{Demers2002} classified V3 (their C05) as carbon rich using narrow-band colours, while \citet{Gullieuszik2007} classified V2 (GRCHH\,24) and V3 (GRCHH\,2) as C-rich on the basis of their near-infrared colours. 
The variables, V1, V2 and V3 are, respectively, about 27,  58 and 104 arcsec from the centre of the galaxy (19:29:59.0 --17:40:41), as can be seen in Fig.~\ref{fig_Varchart}.

In their table 3, \citet{Boyer2015b} list eight suspected variables in Sgr~dIG, including our V2 and V3, as well as GRCHH 14 and 29 which were discussed above. Of the other four, two are faint ($[3.6]>19$) and one of those two is 5.5 arcmin from the centre of the galaxy, while a third suffered from column pull down in the Spitzer imaging; the eighth source is quite faint on the $K_S$ images, is not obviously red and has a close companion. None of these is a strong candidate Mira variable.
V1 is the brightest of six stars that \citet{Boyer2015b} list (their table 4) that are not obviously variable, but which are x-AGB candidates on the basis of $[3.6]-[4.5]$. The first 4 of these correspond to GRCHH 6, 7, 12 and 19. GRCH 19 was discussed above, GRCHH 6, 7 and 12 are relatively blue ($J-K_S\lesssim 1.3$) and only moderately bright ($K_S \gtrsim 17.3$).  None of these 3 is a  highly likely Mira candidates. The fifth star (DUSTiNGS 50071) is a Mira candidate, but can be seen to have a close red companion in the HST images, which will confuse the $K_S$ photometry. 

For the Mira candidates, observations taken within 4 days of each other were averaged to provide a single measurement and periods were determined as in our earlier work \citep{Whitelock2009}. The resulting $JHK_S$ Fourier means (mag), peak-to-peak amplitudes of the Fourier fitted light curves ($\Delta$mag), and periods (P) are listed in Table~\ref{tab_varsP}, where N epochs were used for the quoted data and `redC' is the reddening corrected magnitude. The statistical errors in the period are very small and systematic effects, such as long term variations in the mean light level, suggest a conservative uncertainty of less than 10\%.

\subsection{The M star: V1}\label{Mstar}
\begin{figure}
\includegraphics[width=8.5cm]{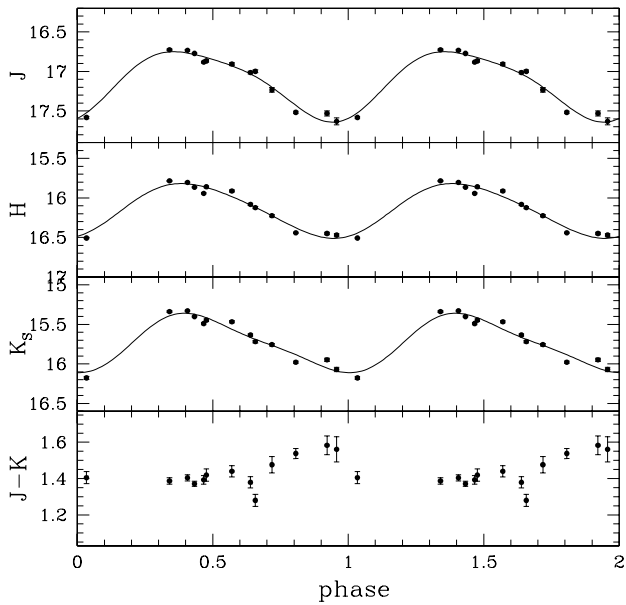}
\caption{$JHK_S$ light curves of V1, arbitrarily phased at a period of 950 day, each point is plotted twice to emphasise the variability.  The solid line show the best fit second-order curve at that period. This plot shows only the data from Table~\ref{tab_Var}.}
\label{fig_Var1}
\end{figure}

The light curves of the M star appear asymmetric (see Fig.~\ref{fig_Var1}), with the rising branch steeper than the falling branch, and for this reason the Fourier fits were made in the second order. This should not be over-interpreted as the observations clearly cover two falling branches, but we have no measurements over the time period during which it brightens. 

Doing the period analysis from the data in Table~\ref{tab_Var} together with the additional $H$ and $K_S$ photometry from Table~\ref{tab_extraV} results in periods for V1 of 946 days at $H$ and 1055 days at $K$. These values are strongly influenced by the isolated observations, 
so we can be confident that this star has a period of the order of 1000 days, but a much longer time series is required to establish the exact value.  For the discussion below we use a period of 950 days derived from the IRSF  $K_S$ data.

\begin{table}
\centering
\caption{The variable stars. (G is the GRCHH number and Dusti is the DUSTiNGS number from \citet{Boyer2015a}.}
\begin{tabular}{cccccccc}
name & G &RA & Dec & Dusti & [3.6] & [4.5] & sp\\ 
& &\multicolumn{2}{c}{(2000) }\\
V1  &21 &19:29:57.9& --17:40:17 & 44334 & 14.83 & 14.17 & M\\
V2 &24 &19:29:56.1 & --17:41:21 & 47717 & 14.89 & 14.13 & C\\
V3 &2 &19:30:06.3 & --17:40:55 & 29075 & 14.34 & 13.84 & C\\
\end{tabular}\label{tab_var2}
\end{table}

\begin{figure}
\includegraphics[width=8.5cm, bb=0 200 574 550]{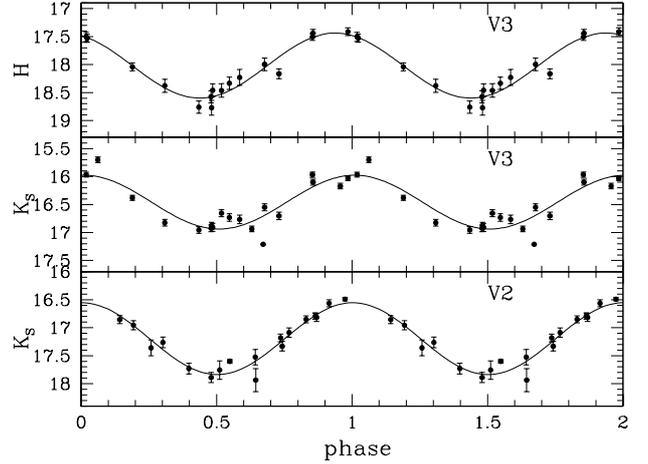} 
\caption{The $K_S$ light curves of V2, arbitrarily phased at a period of 670 days and the $H$ and $K_S$ light curves of V3, phased at 503 days.  Each point is plotted twice to emphasise the variability and the solid lines show the best fitting sine curve at the given period.} 
\label{fig_Var2}
\end{figure}

\begin{figure*}
\includegraphics[width=2.8cm]{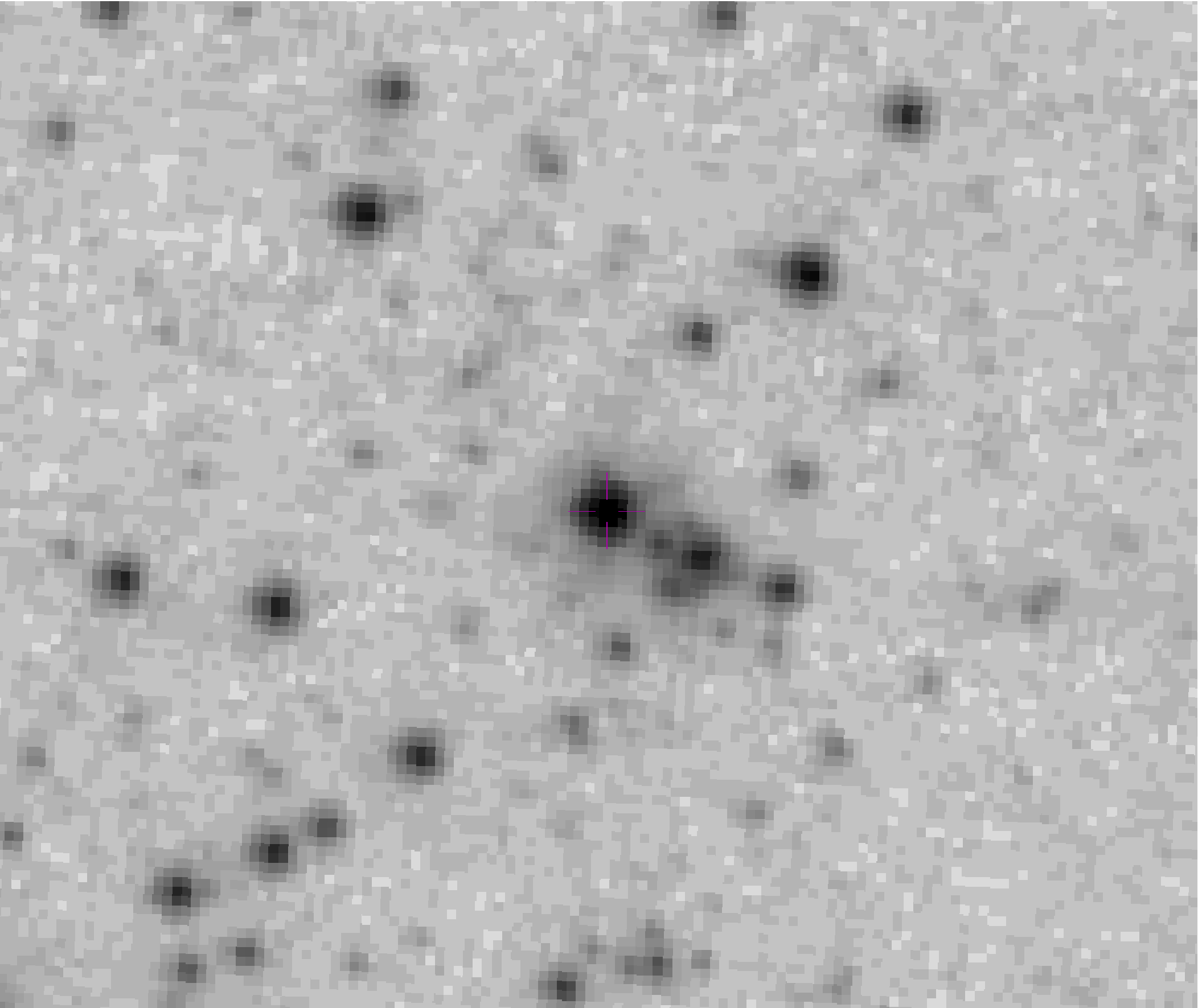}
\includegraphics[width=2.8cm]{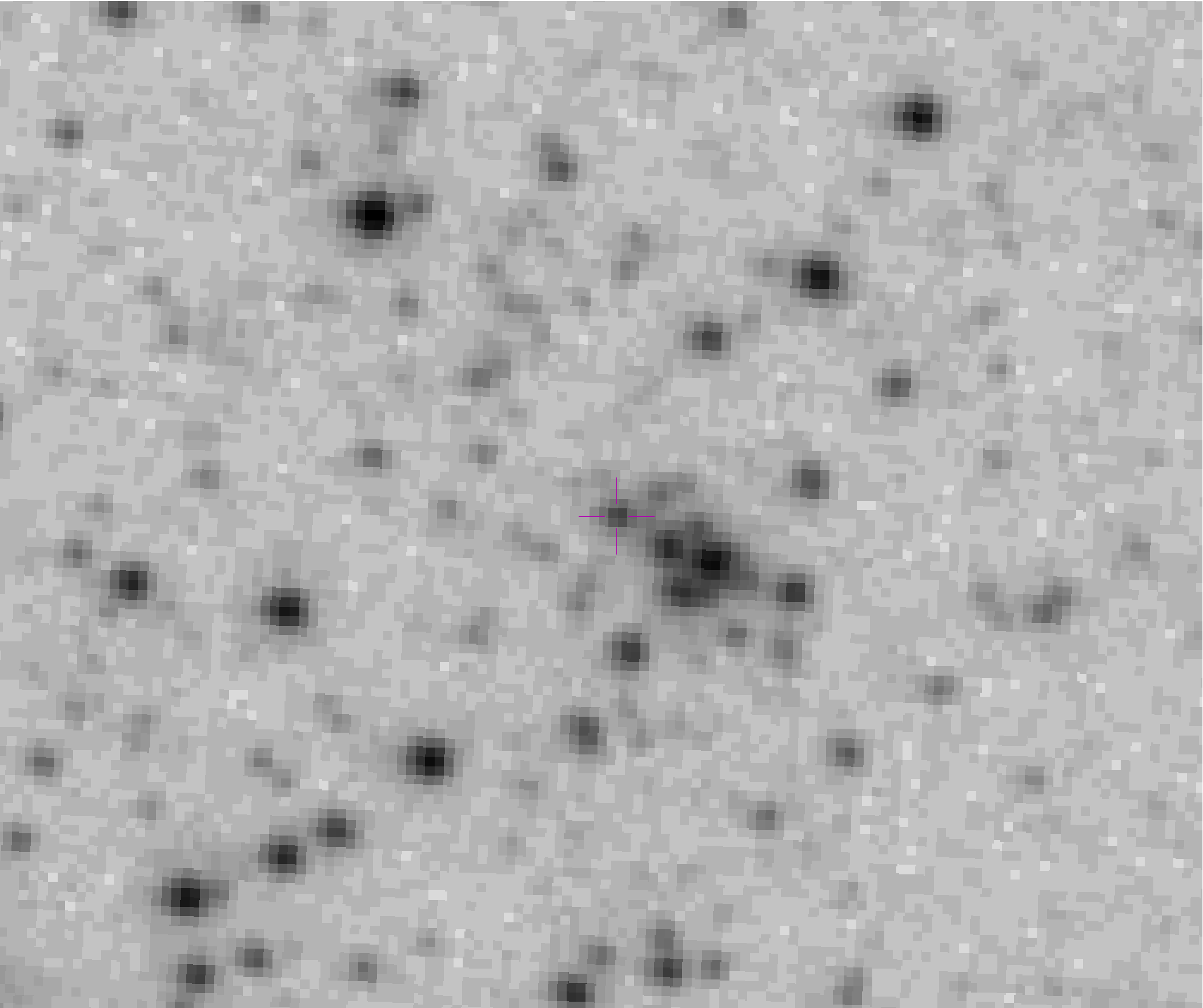}
\includegraphics[width=2.8cm]{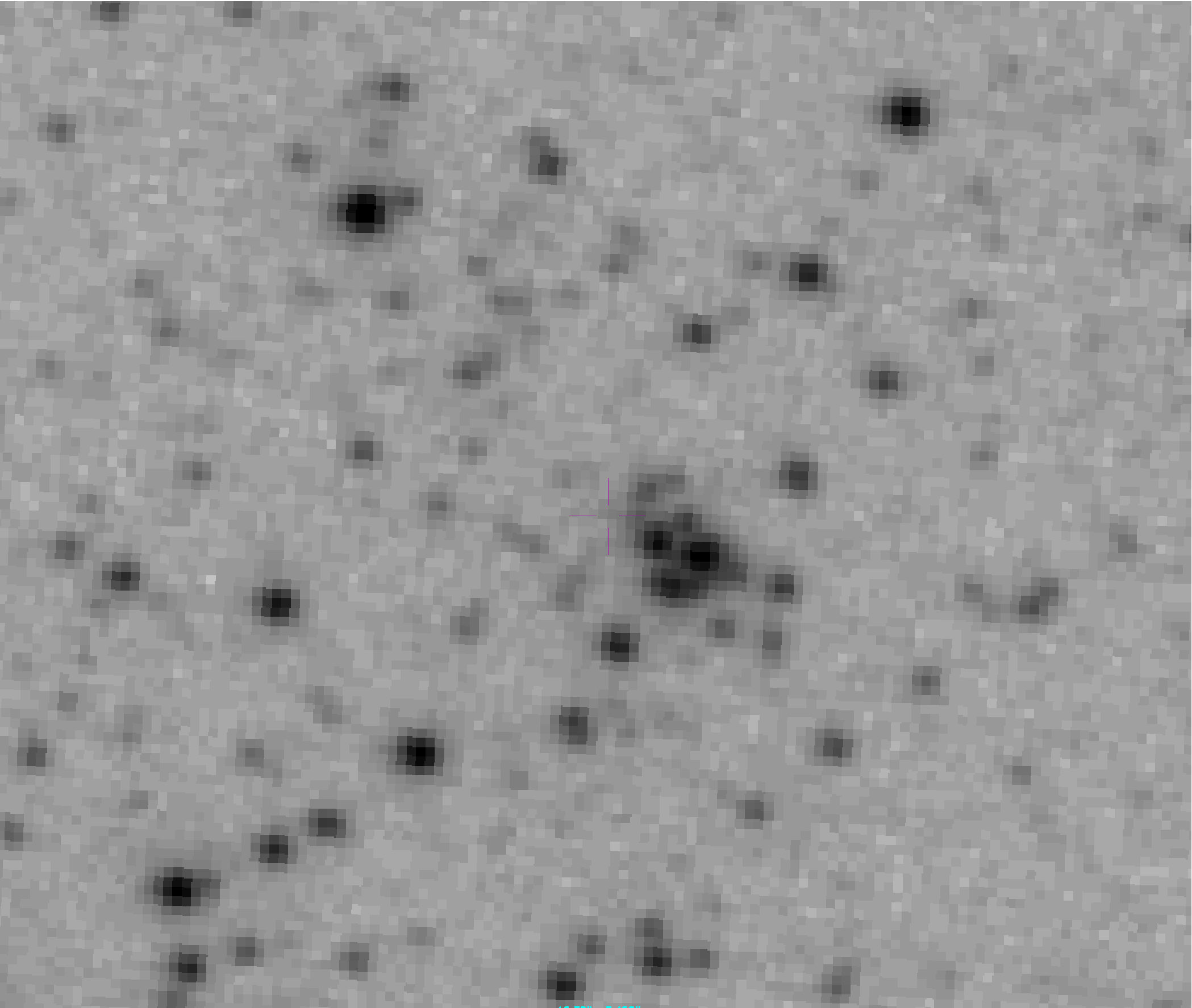}
\caption {Comparison of HST 814, 606 and 475nm images (left to right) centred on the variable star. The field is about 5 arcsec square with N up and E to the left.}
\label{fig_stamps}
\end{figure*}

 As noted above, \citet{Gullieuszik2007} find V1 to be blended. It is vital to establish to what extent this is the case if we are to interpret our photometry correctly. Figure \ref{fig_stamps} shows the region around V1 as seen in the 2003 HST archival images in the [814], [606] and [475] bands. 

In the \citet{Momany2014} catalogue, there are measurements in 3 bands for 11 stars inside a circle of diameter 3 arcsec centred on the variable. Using these data we estimate that the neighbouring stars contribute only about 0.1 mag to the $J$ mean magnitude of the variable, and probably somewhat less to $K_S$, so its position in the $K_S$, $J-K_S$ (Fig.~\ref{fig_cm}) diagram is little affected by these stars. Curiously, the catalogue contains no photometry for V1. We have  carried out aperture photometry on the star on the HST frames and find [475], [606] and [814] magnitudes of $\ge28.0$, 24.34 and 20.62, respectively.

The fact that the amplitude (Table \ref{tab_varsP}) at $J$ is larger than at $K_S$ and that $J-K_S$ is reddest when $K_S$ is faintest (Fig.~\ref{fig_Var1}) both support the conclusion that the photometry is not confused by any nearby bluer source. 

For stars of this colour the correction from the 2MASS system to the SAAO system\citep{Carpenter2001}\footnote{update at:$\rm www.ipac.caltech.edu/2mass/releases/allsky/doc/sec6 \_4b.html$}  is essentially zero, so correcting only for reddening we have $K_0=15.7$.  The  PL relation from \cite{Whitelock2008} (with LMC $(m-M)_0=18.5$),  $M_K=-3.51\log P +1.09$, would give $M_K=-9.36$. So at the distance of Sgr~dIG we should expect $K_0=15.84$. This agreement is surprising and possibly coincidental, unless this star has recently stopped HBB as suggested below (see  sections~\ref{discus} and \ref{theory}).

\begin{table*}
\centering
\caption{Derived data for the variables, redC lists the reddening corrected mags, while $m_{bol1}$ and $m_{bol2}$ are derived from the colour-dependent bolometric correction and from integrating the flux, respectively.}
\begin{tabular}{cccccccccc}
      &  mag   & $\Delta$mag & N  & Period & redC   &$m_{bol1}$ & $m_{bol2}$     & $M_{bol}$     \\
\multicolumn{4}{l}{V1}            & 950    &        &  --       & $18.50\pm0.15$ & $-6.70\pm0.2$ \\
$J$   &  17.14 &  0.90       & 13 & 914    & 17.04  \\
$H$   &  16.15 &  0.70       & 13 & 933    & 16.09\\
$K_S$ &  15.74 &   0.75      & 13 & 953    & 15.70 \\
\multicolumn{6}{l}{V2}                              & 19.45     & $19.53\pm0.20$ & $-5.67\pm0.25$ \\
$K_S$ &  17.18 &   1.21      & 15 & 670    & 17.14 \\
\multicolumn{6}{l}{V3 WISE}                    & 19.86      & $19.46\pm0.2$ &  $-5.74\pm0.25$\\
\multicolumn{7}{l}{V3 Spitzer}                                  & $19.01\pm0.10$&  $-6.19\pm0.15$\\
$H$   &  18.02 &   1.16      & 16 & 504    & 17.96 \\
$K_S$ &  16.46 &   0.96      & 19 & 503    & 16.47 \\
\end{tabular}\label{tab_varsP}
\end{table*}

\subsection{The C stars: V2, V3}\label{Cstar}

For V2 we only have the $K_S$ time series, combining the data from Tables  \ref{tab_extraV} and \ref{tab_Var}, and the colours from \citet{Gullieuszik2007}, as the star was too faint for us to measure at shorter wavelengths. The same applies to V3 although we also have $H$ magnitudes for this bluer star. It is possible to estimate the bolometric magnitudes for these stars using our Fourier mean $K_S$ magnitudes and the $(J-K_S)$ colours from \citet{Gullieuszik2007}, noting that they observed V2 close to  minimum and V3 close to maximum light and that their photometry is on the LCO system \citep[see][]{Carpenter2001}. $K_S$ and $(J-K_S)$ are converted to $K$ and $(J-K)$ on the SAAO system following Carpenter (see section \ref{Mstar}), and then used to estimate the bolometric correction and magnitude following  \citet{Whitelock2006}.  These can be compared to the bolometric magnitudes predicted by the PL relation (equation A1, $M_{bol}=3.89-3.31\log P$, from \cite{Whitelock2009}, after correcting the LMC distance modulus from 18.39 to 18.5), noting the caveats in section \ref{Mira_PLR}. 

V2 (GRCHH 24) is the reddest of the stars in table 4 of \citet{Gullieuszik2007}, for which they quote  $J-K_S=4.147$. We use $J-K_S \sim 4.0$  as the estimated mean 2MASS colour and derive SAAO-system reddening-corrected values of $K_0=17.13$ and $(J-K)_0=4.3$, leading to $m_{bol}=19.45$ and $M_{bol}=-5.75$, compared to $M_{bol}=-5.46$ predicted by the PL relation.
 
V3 (GRCHH 2) has $J-K_S=2.360$ in table 4 of \citet{Gullieuszik2007}. We use $J-K_S \sim 2.4$ as its mean 2MASS value and derive $K_0=16.4$ and $(J-K)_0=2.5$ on the SAAO system, leading to $m_{bol}=19.86$ and  $M_{bol}=-5.34$, compared to $M_{bol}=-5.05$ predicted by the PL relation.

Thus both of the C stars are about 0.3 mag brighter than the linear PLR would predict. Given the uncertainty in the colour, the bolometric correction and the distance this difference is not significant. 
It seems likely that Sgr~dIG V2 and V3 represent the longest period and most luminous examples of the C-rich Miras in Sgr~dIG and their initial mass is discussed in section \ref{theory}. Short period Miras may also be present. 

\subsection{Flux Calculation and Bolometric Magnitudes}
\begin{figure}
\includegraphics[width=8.5cm]{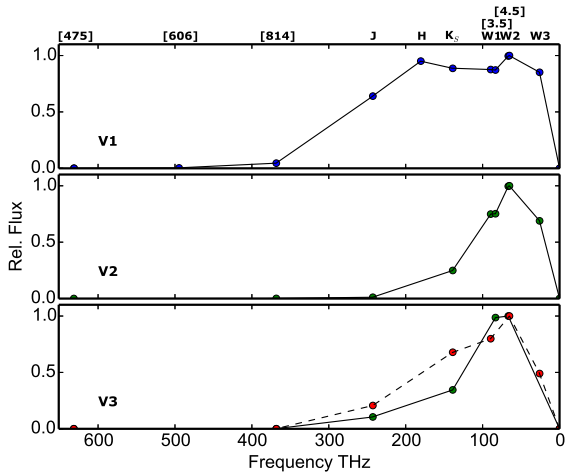}
\caption{Flux curves for V1, V2 and V3 in Sag dIG, normalised to maximum flux (usually in the W2 band). The continuous lines show a linear interpolation between points. For V3, the continuous lines connect the near-infrared points and the Spitzer points, while the dashed line connects the near-infrared points and the WISE points - see text for more details. The bands to which the data points refer are shown above the top plot.}
\label{fig_flux}
\end{figure}

We have made use of all available photometry to construct flux curves for  the variables. Apart from the IRSF $JHK_S$ measurements reported above, we have measured V1 and V2 on the HST images referred to above to give [475], [606] and [814] magnitudes for V1 and an [814] magnitude for V2. V3 is outside the field of the HST images. The $K_S$ magnitude for V2 was found from our photometry of the ESO images referred to above. Mid-infrared magnitudes were obtained from WISE photometry \citep{Cutri2012}, and Spitzer [3.6] and [4.5] magnitudes from \citet{Boyer2015a, Boyer2015b}. The photometry was corrected for reddening using the curve from \citet{Schlafly2016} adjusted to give $A_V/E_{(B-V)}=3.23$. Conversion from magnitude to flux was carried out using zero-point data obtained from the Gemini web page (http://www.gemini.edu/sciops/instruments). Integration under the curves resulted in apparent bolometric magnitudes. Errors are determined on the basis of the catalogued WISE and Spitzer values. 
Further uncertainty, which is difficult to quantify, is introduced because the photometry longward of $2.2 \mu$m refers to different phases and light curves are not available in those wavebands to allow the appropriate magnitudes to be used. The amplitudes of these dusty C stars at [3.5] and [4.6] are typically one magnitude, but can be more \citep[see][fig.~1]{Whitelock2017}. The normalised energy distributions of the variables are shown in Fig.~\ref{fig_flux} and the bolometric magnitudes are given in Table~\ref{tab_varsP}. 

This simple method of estimating the integrated flux gives a bolometric magnitude ($-5.7$, 19.53) for V2 that is close to the value derived in  section \ref{Cstar} from the bolometric correction ($-5.7$, 19.45).   These values are also close to the expectation from the PL relation for C stars ($-5.46$, 19.74;  see section~\ref{Cstar} and \citet{Whitelock2009}). 

In the case of V3, the catalogued value of W3 is listed as an upper limit, so we estimated W3 from fig.~19 of \citet{Nikutta2014}, assuming an error of $\pm0.3$. The flux derived from the near-infrared and WISE data  ($-5.7$, 19.46) is 0.4 mag brighter than that derived from the bolometric correction ($-5.3$, 19.86). Inexplicably, the Spitzer [3.6] and [4.5] magnitudes are about 1 mag and 0.7 mag brighter than the WISE W1 and W2 values; the mean phases of the two data sets are about the same, though the Spitzer data were obtained about one cycle later than the WISE ones. 


For V1, we find a bolometric magnitude of (-6.70, 18.50) from the flux integration. We do not have a calibration of  bolometric correction with colour for stars with such unusual flux distributions. Its luminosity is discussed further in the next section.

\begin{table*}
\centering
\caption{Colours, amplitudes and periods of HBB candidates in IC\,1613  ($[3.6],[4.5]$ are multi-epoch mean values from the SPIRITS survey \citep{Kasliwal2017}; other data are from \citet{Menzies2015}) and Sgr~dIG.}
\begin{tabular}{cccccccc}
name & P &  $\Delta K$ &$M_K$ & $M_{[3.6]}$  & $(J-K_S)_0$ & $J_0-[3.6]$ & $ [3.6]-[4.5]$\\
3011 & 550  & 0.82 &-9.20 & -9.61 & 1.01 & 1.42 & 0.18 \\
2035 &  530 & 0.75 & -9.17 & -9.66 &1.13 & 1.62 &  0.09 \\
1016 &  464 & 0.50 &-9.33 & -9.46 & 1.14 & 1.27 & 0.09 \\
V1 &  950 & 0.75 & -9.5 &  -10.37 & 1.34 & 2.21 & 0.66\\
\end{tabular}\label{HBBstars}
\end{table*}

\section{Discussion of V1 and the PLR}\label{discus}


Given the unusual characteristics of this star as described by \citet{Boyer2015a} and Boyer et al. (2017 in preparation), i.e. extreme AGB mid-infrared colours and magnitudes and the $\rm  H_2O$ absorption characteristic of an M star, we reject any possibility that it is not the same star as that for which we have determined the period. A comparison with the colour-magnitude diagram for IC\,1613 \citep{Menzies2015} and isochrones from \citet{Marigo2008} shows V1 to be where HBB variables are expected. Its position in the galaxy, within the volume occupied by young blue stars, is consistent with its being relatively massive and the comparison with theory in section \ref{theory} provides more insight.
V1 is slightly brighter and redder than the IC\,1613 stars in the near-infrared, but considerably redder in $[3.6]-[4.5]$ (see Table~\ref{HBBstars}).

\begin{table*}
\centering
\caption{V1 and various PLRs (columns: (2) reddening corrected mean magnitude (3,4,5) PLR predictions from: (3) \citet{Yuan2017}, (4) \citet{Ita2011}, (5) \citet{Whitelock2008, Whitelock2009}, \citet{Riebel2015}, (6,7,8) differences between observed and predicted values).}
\begin{tabular}{lcccccccc}
(1) & (2) & (3) & (4) & (5) & (6) & (7) & (8)\\
band &  redC & Yuan+ &  I+M  & S/R & \multicolumn{3}{c}{observed--PLR}\\
 &&  &&&Yuan+ & I+M& S/R \\
 \multicolumn{3}{l}{Sgr dIG}\\
$K/K_S$ & 15.70 & 14.43 & 14.53 & 15.84 &  1.27 & 1.17 & --0.14\\
$[3.6]$        & 14.83 & 13.92 & 14.22 & 15.03  &  0.91& 0.61 & --0.20\\
$[4.5] $       & 14.17 & 13.77 & 14.14 & 15.16  & 0.40 & 0.03 & --0.99\\
$m_{bol} $    & 18.50 &  -       &18.00   & 19.23 & -        & 0.50 & --0.73\\
\multicolumn{3}{l}{IC\,1613 \#3011}\\
$K/K_S$ & 15.17 & 15.22 & 15.32 & 15.85 & -0.05 & -0.15 & -0.67 \\
$[3.6]$    &  14.76 & 14.77 & 14.72 & 15.15 & -0.01 & 0.04 & -0.39 \\ 
$[4.5] $    & 14.58  & 14.66 & 14.56 & 15.23 & -0.08 & 0.02 & -0.65 \\
$m_{bol} $& 18.33 &   -       & 18.19  & 19.19 &  -      & 0.14 & -0.86 \\
\end{tabular}\label{PLRcomp}
\end{table*}

\begin{figure}
\includegraphics[width=8.5cm]{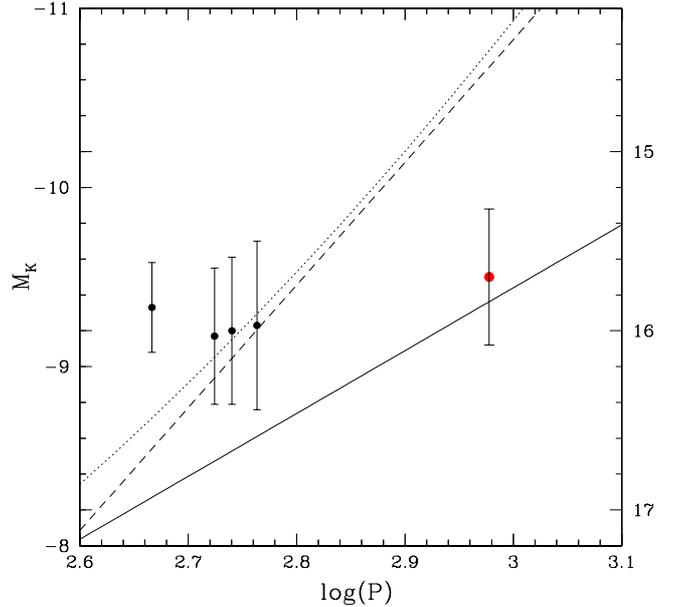}
\caption {The $K$ PLR showing V1 (red) and the four HBB stars (black) in IC\,1613 from \citet{Menzies2015}. The solid, dashed and doted lines show, respectively, the PLRs from \citet{Whitelock2008}(an extrapolation), \citet{Ita2011} and \citet{Yuan2017}; all relations are for O-rich stars. The error bars show the range of variability. } \label{plk}
\end{figure}

The long period of V1 puts it in a part of the PLR where HBB is important and the extrapolated SAAO relation \citep{Whitelock2008} falls significantly below those from \citet{Ita2011} and \citet{Yuan2017}(see section~\ref{Mira_PLR}).
 This is illustrated in Fig.~\ref{plk} which shows the various $K_S$ PLRs together with the data for V1 and the four HBB stars from IC\,1613.  The error bars show the amplitudes of the sine curves that best fit the $K_S$ light curves, and are therefore slightly smaller than the observed spread (PLRs for Miras are sometimes shown with single observations rather than mean magnitudes and it is important to appreciate the spread that arises simply from variability). Three of the four HBB Miras in IC\,1613 lie close to the bright PLRs, while the fourth is even brighter (and could possibly be an overtone pulsator).  V1 is about 1.2 mag fainter than the bright PLRs and on the extrapolated SAAO PLR. 
	
The Spitzer photometry of V1 is the mean of two epochs (JD\,2455886 and JD\,2456089)  \citep{Boyer2015a} and it was not recognised as variable. If the mid-infrared light-curve shows the same behaviour as the near-infrared one, both the DUSTiNGS observations were obtained at about the same phase, corresponding approximately to mean light. Table~\ref{PLRcomp} lists the deviation of V1 from the various PLRs. It is brighter at [3.6] and [4.5] than the \cite{Riebel2015} relation (columns 5 and 8), which is a single straight line over the entire period range. Although it is slightly (0.4 mag) fainter than the \citeauthor{Yuan2017} PLR (columns 3 and 6), it is very close to the \citeauthor{Ita2011} (columns 4 and 7) relation at [4.5].

\citet{Whitelock2017}  discuss the [3.6] and [4.5] PLR relations for LMC and IC\,1613 AGB variables, noting that, for C stars, there is a spread at long periods which is a function of colour. Their sample of long period O-rich stars was rather small.
It is informative to note the position of V1 in these PLR plots  \citep[][figs.~2 and 3]{Whitelock2017} where, at the distance of the LMC, it would have [3.6]=8.1,  [4.5]= 7.5 and $K=9.0$; note that V1's period is longer than that of any variable discussed by \citeauthor{Whitelock2017} At [3.6] it is not far from the extrapolated \citet{Riebel2015} relation, while at [4.5] it is considerably brighter. 
It can also be compared with \citeauthor{Whitelock2017} fig~4 which shows AGB variables in IC\,1613, including the HBB stars discussed here, in a similar PL diagram.  Clearly the PLR for long period variables is complex and requires further study.
	
We can compare the bolometric magnitude given in Table~\ref{tab_varsP} with the predictions of the \citet{Ita2011} PLR (columns 4 and 7) and linear extrapolation of the \citet{Whitelock2009} PLR (columns 5 and 8).  It lies between the two relations and at $M_{bol}\sim-6.7$ is about half a magnitude brighter than the 4 HBB stars in IC\,1613, which have $M_{bol}$ between --6.0 and --6.3. This is to be expected since the flux distribution (Fig.~\ref{fig_flux}) shows that the bolometric magnitude will depend strongly on the mid-infrared flux. 

Table~\ref{PLRcomp}, for comparison, tabulates  the deviation from the PLRs of one of the IC\,1613 HBB stars \citep[no. 3011 from][]{Menzies2015}. This star falls very close to both the \citet{Yuan2017} and \citet{Ita2011} relations, at all wavelengths (see columns 6 and 7). 

The appendix (section \ref{appendix}) to this paper describes a search for stars similar to V1 in the LMC (i.e. long periods, blue colours and low $K_S$ luminosities), but no convincing candidates are identified. Thus it seems that although V1 has many of the characteristics of a HBB Mira it is distinctly different from such stars known in other galaxies. 

\section{Comparison with theoretical models}\label{theory}
\begin{figure}
\includegraphics[width=8.5cm]{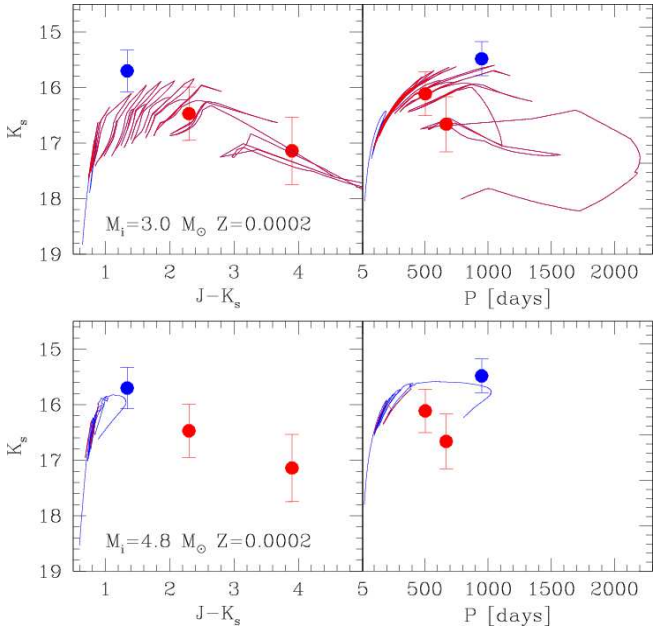}
\caption {Evolutionary tracks in  colour-magnitude and period-luminosity diagrams
(C stars in red and V1 in blue).
Error bars show the variability range. AGB evolutionary tracks are shown for two choices of the initial mass as indicated and metallicity Z=0.0002. Stages characterised by surface $\rm C/O<1$  and $\rm C/O>1$ are coloured in blue and red, respectively. } \label{tracks}
\end{figure}

Thermally Pulsing (TP)-AGB tracks are computed with the {\sc colibri} code \citep{Marigo2013} for  two values of the initial stellar mass ($3.0 \, M_{\odot}$ and $4.8 \, M_{\odot}$, see Fig.~\ref{tracks}) and initial composition with metallicity Z=0.0002, and helium abundance Y=0.249. This choice is suitable to represent the abundance ratio $\rm [Fe/H] = -1.88$, assuming a metallicity for the Sun of $Z=0.0152$ \citep{Caffau2011,Bressan2012}. The evolution is followed up to the complete ejection of the envelope by stellar winds. During the initial stages on the Early-AGB  mass loss is described with a modified version of the \citet{Cranmer2011} formalism  suitable for cold chromospheres.  At larger luminosities, when radiation on dust is expected to drive the wind, we adopt the \citet{Bloecker1995} formula (with the efficiency parameter $\eta=0.05$) as long as the surface $\rm C/O<1$, and the routine provided by \citet{Mattsson2010} for $\rm C/O>1$, based on dynamic atmosphere models for pulsating C stars. More details of the models will be provided in a forthcoming theoretical paper (Pastorelli et al., in preparation).

TP-AGB models account for the changes in the envelope composition due to the occurrence of the third dredge-up and HBB. The effect of light reprocessing by circumstellar dust in the extended  envelopes  of  mass-losing  AGB stars is included following the approach of \citet{Marigo2017}. For C stars, in particular, we use  the most recent updates of the dust-growth model and radiative transfer calculations presented by \citet{Nanni2016}.  For models with $\rm C/O < 1$ we used the tables of bolometric corrections calculated by \citet{Marigo2008}. These are based on radiative transfer models presented by \citet{Bressan1998}. Pulsation periods for the fundamental mode  are computed as a function of stellar parameters  with the aid of the analytic fits in \citet{Marigo2017}, based on new models for long-period variables (Trabucchi et al., in preparation).

The $3\, M_{\odot}$ model fits well with the observed photometric and pulsation properties of the two C stars, providing  support for the calibration of the dust properties carried out by \citet{Nanni2016}. The current masses  for the two C stars should be in the range $1.5-1.9\,  M_{\odot}$.

The M star is interpreted as an AGB star of initial high mass ($\sim 4.2 - 4.8 \, M_{\odot}$) which has a relatively low current mass ($\sim 1.1 -1.4 \, M_{\odot}$) in a phase of intense mass loss, soon after the extinction of HBB. The third dredge-up is also quenched due to the very low residual envelope mass ($0.2-0.4\, M_{\odot}$), so that the formation of a late C star is prevented. We recall that the adopted \citet{Bloecker1995} formula has a strong dependence on the  luminosity ($\propto L^{4.25}$). As a consequence, it predicts high mass-loss rates in bright low-metallicity HBB stars despite their relatively high effective temperatures.  According to our present set of stellar models, the initial mass of the progenitor cannot be lower than $\sim 4.2\, M_{\odot}$ since below this mass limit TP-AGB stars are predicted to end their lives as C stars and they do not reach the high luminosity measured for V1.     The model with $M_{\rm i} =4.8 \, M_{\odot}$ represents the upper limit for a star with $Z=0.0002$ to develop a degenerate C-O core and experience the AGB phase. Just above this mass limit we expect to have stars that enter the super-AGB phase after the end of the C-burning phase.   We cannot exclude the possibility that V1 may be a super-AGB star, in which case its luminosity would suggest it is near the lower limit for such stars.

 The rather blue $JHK_S$ colours of V1 are likely the result of the low abundance of silicon (the key-element for the formation of silicate grains) which limits the amount of silicate dust formed, as well as the relatively high effective temperatures of the models (4500-4000 K) which move the dust condensation radius to larger distances from the star. All this makes dust formation and reddening less efficient. These aspects need a careful investigation through consistent dynamical atmospheres models \citep[e.g.][]{Bladh2015}.

The long period of 950 days is only reached during the final AGB stages when the mass has been significantly reduced by stellar winds. So the star will soon leave the AGB and enter the post-AGB phase. The time-scales involved are all very short. Our calculations predict that the duration of the TP-AGB phase (from the first thermal pulse up to envelope ejection) of stars in the relevant initial mass interval (from $4.2 - 4.8 \, M_{\odot}$) ranges from $\sim 7\times 10^4$ yr to $\sim 2.5\times 10^4$ yr. The last interpulse period during which the mass-losing stars quickly accelerate towards very long periods (from $\sim 400$ days to $\sim 1000$ days) and then leave the AGB is of the order of a few $10^3$ yr (from $\sim 9.7\times 10^3$ yr to $\sim 3.6\times 10^3$ yr).

\section{Cepheids in Sgr dIG}

As we are suggesting that V1 is massive in AGB terms ($> 4 \, M_{\odot}$, see section \ref{theory}) and given that short period Cepheids presumably evolve into long period Miras, it is worth considering if Cepheids should also be detectable in Sgr~dIG.  

With the IRSF we would have detected any large amplitude variables with mean $K_S<17.2$. A Cepheid with $K_S=17.2$ would have a period of 50 days \citep{Matsunaga2013} and a mass about $9 \, M_\odot$ \citep{Anderson2016}, i.e. considerably larger than is likely for V1.  Such long period Cepheids are rare; there are only two in NGC\,6822 with P$>$50 days and none in IC\,1613, and both these galaxies are significantly more massive than Sgr~dIG (as discussed in section \ref{LG_comp}).  We could well expect Sgr~dIG to contain  Cepheids with periods of a few days, but these are too faint to detect with a 1.4-m telescope (a Cepheid with P=3 days would have $K_S\sim 21.5$).

This is an illustration of the potential importance of Mira variables in the era of JWST and large ground-based telescopes optimised at infrared wavelengths. Miras will be easily detected where individual stars of low or intermediate mass can be resolved. They should prove very useful probes of stellar populations as well as for the distance scale \citep{Feast2014, Whitelock2014}.






\begin{table*}
\centering
\caption{Comparison of Mira variables in Local Group dIGs}
\begin{tabular}{lccccccc}
name & M& NO$_{HBB}$ &   P range & NC & P range &  $N_{\rm other}$ \\
 & ($10^{6} M_{\odot}$) && (day) & & (day)\\
NGC\,6822 &  100 & 4(5) & 545-638 (854) & 50 & 182-998 & 27 \\ 
IC\,1613     &  100 & 4 &  464-580 &  9 & 263:-430(879) & $>9$ \\
Sgr dIG      &  3.5  & 1 & 950 & 2 & 503-670 & 2 \\
\end{tabular}\label{digs}
\end{table*}

\section{Comparison with other dwarf Irregular galaxies}\label{LG_comp}
AGB variables in NGC\,6822 and IC\,1613 were discussed by \citet{Whitelock2013} and\ \citet{Menzies2015}, respectively. The numbers of large amplitude variables from those galaxies are compared with the numbers found in Sgr~dIG in Table~\ref{digs}, where M is the total visible mass of stars in the galaxy \citep[from][]{McConnachie2012},   NO$_{HBB}$ is the number of (presumed) O-rich Miras with periods over 450 days, i.e., those we can be reasonably certain are undergoing HBB, P range (col 4)  is their range of periods, NC is the number of (presumed) C rich Miras followed by their period range (col 6), while N$_{other}$ is the number of other large amplitude variables found for which no periods could be determined.

The survey of NGC\,6822 is the most nearly complete, although even there a few C-rich Miras were missed (as discussed in section 6 of that paper) and there may be a very small number of Miras outside the area surveyed. The survey of IC\,1613 did not cover the full area of the galaxy and will be incomplete. We can estimate that the total number of C-rich Miras in the areas surveyed will be given by the sum of NC+$\rm N_{other}$, i.e. 77 and 18 for NGC\,6822 and IC\,1633, respectively. We would not necessarily expect to find very long period Miras in these galaxies, but none of the surveys would have found very red stars with $P>1000$ day. It is also quite possible that some short period O-rich Miras, of the type found in Galactic globular clusters \citep[e.g.,][]{Feast2002}, will have been missed in all three galaxies.

Because the long period O-rich (HBB) Miras are from relatively massive progenitors they will be confined to the central regions of their respective galaxies. This, together with their brightness, colour and large amplitude variations, suggests that the count of these stars should be complete for all three galaxies. The number for NGC\,6822 is given as either 4 or 5 because of the uncertain status of the longest period star in that galaxy.  This may be a supergiant rather than an AGB star and its low amplitude ($\Delta K_S \sim 0.4$) would support that interpretation. If it is an AGB star then it is very interesting and a candidate super-AGB star, with $M_K= -10.9$  and $M_{bol}\sim-8.0$; it is certainly worth further investigation. NGC\,6822 also has six lower luminosity O-rich Miras with periods between 158 and 402 days, which may be similar to globular cluster Miras.

Thus for Sgr~dIG we can be certain that V1 is the only HBB Mira in the galaxy and a remarkable star it is. If we normalise the HBB variable count by the relative masses of NGC\,6822, IC\,1613 and Sgr~dIG (Table~\ref{digs}) we expect, as we do indeed find, the same number in NGC\,6822 and IC\,1613. In Sgr~dIG we expect  0.04, i.e. none at all.  The galaxy masses are not accurate, but the existence of this star is nevertheless quite striking. As noted in section~\ref{intro},  Sgr~dIG has a particularly high specific star formation rate. It seems likely that V1 is an evolved AGB star with a relatively massive progenitor (section \ref{theory}), but the situation is complex and beyond the scope of this discussion.

Turning now to the carbon stars, we normalise by the relative mass of NGC\,6822 and Sgr~dIG and expect to find  2 or 3  C-rich Miras in Sgr~dIG, where we actually find 2 definite plus 2 probable (there is at least one more possible candidate from \citet{Boyer2015b} as discussed in section~\ref{var_stars}). This may suggest that our survey of C-rich Miras is very nearly complete.  Although it is more likely that the star formation history is more complex than our simple comparisons assume. There are in any case many reasons for investigating Sgr~dIG in more detail. 

\section{conclusion}

Two large amplitude carbon Miras have been identified in Sgr~dIG and, within the limits of our investigation, these are comparable to similar stars in other Local Group galaxies  and they probably evolved from stars with main sequence masses $M_i \sim 3\, M_{\odot}$.

The single O-rich Mira has a particularly long period  (950 days), and must have been undergoing HBB.  Its observed characteristics are different from all other well-studied AGB variables. Although its mid-infrared colours suggest a dust excess, its near infrared colours are rather blue.  Its mean magnitudes do not tie in with the expectations of the PLRs which include HBB stars \citep{Ita2011,Yuan2017} (perhaps because it has recently stopped HBB); in particular, it is faint at $K_S$ and bolometrically, although close to the \citeauthor{Ita2011} PLR at [4.5].  It would be useful to know more about its longer wavelength flux.

 A comparison with models suggests that V1 started with a main sequence mass of around $4.8 \, M_{\odot}$ and is now in a very advanced short-lived evolutionary state. The fact that we have only found a single star with these characteristics in combined surveys of the LMC, NGC\,6822, IC\,1613 and Sgr~dIG makes a short-lived state plausible, but other alternatives should be considered. It is  possible that V1 is in an interacting binary system. Binary Miras are not so unusual and symbiotic Miras have distinct differences from apparently solitary ones\citep{Whitelock1987, Mohamed2012}. In any case this star is  well worth further  study, and monitoring for the changes which might be anticipated if it is in a short lived evolutionary state. 

We suggest that the existence of this star and its unusual characteristics may be related to the nature of Sgr~dIG, perhaps linked to its very high specific star formation rate, which increases the probability of finding a relatively massive AGB variable despite the low total stellar mass. The low metallicity of Sgr~dIG is also significant; it is (along with Leo A ) the most oxygen-poor galaxy in the Local Group \citep{Kirby2017}.  The fact that the C-rich Miras are similar to those in other Local Group galaxies, but that the O-rich star is different, points to abundance as a factor. The low metallicity and low oxygen abundance will affect the efficiency of dust production and type of dust grains that can form in the atmospheres of O-rich AGB stars. Abundances make much less difference to mass loss in C-rich stars as their dust is formed from dredged up carbon, and dredge-up is efficient at low metallicity \citep{Karakas2002}.  Evidence to date \citep[][and references therein]{Sloan2012} indicates that  in O-rich stars dust production is a function of metallicity, but in C stars it is not.  The relative transparency of O-rich dust compared to C-rich dust in normal Galactic Miras has been discussed by \citet{Bladh2015} and \citet{Hofner2016}, but it is not clear what the O-rich dust will be at very low abundances. Mass-loss is driven through radiation pressure on dust and thus mass-loss and opacities in Sgr~dIG O-rich Miras will be very different from those in better studied AGB stars in the Galaxy and Magellanic Clouds. 

\section*{acknowledgments} MWF, JWM and PAW are grateful to the National Research Foundation of South Africa for research grants. PM acknowledges the support from the ERC Consolidator grant, project STARKEY (G.A. 615604).
This paper made extensive use of the CDS services,  Aladin and VizieR.  It also makes use of the following: data products from the Wide-field Infrared Survey Explorer (WISE), which is a joint project of the University of California, Los Angeles, and the Jet Propulsion Laboratory/California Institute of Technology, funded by the National Aeronautics and Space Administration (NASA);
observations made with the NASA/ESA Hubble Space Telescope, obtained from the data archive at the Space Telescope Science Institute (STScI is operated by the Association of Universities for Research in Astronomy, Inc. under NASA contract NAS 5-26555); observations collected at the European Organisation for Astronomical Research in the Southern Hemisphere under ESO programmes: 61.E-0273(A),71.D-0560(A),60.A-9205(A),59.A-9004(D),077.D-0423(B),079.D-0482(A) and 60.A-9700(E); data obtained from the ESO Science Archive Facility under request numbers 284193, 284208, 284212, 284379 ,284404, 284405, 284415, 286398, 286459, 286588, 286602 and 286607.
The IRSF project is a collaboration between Nagoya University and the SAAO supported by the Grants-in-Aid for Scientific Research on Priority Areas (A) (no. 10147207 and no. 10147214) and Optical \& Near-Infrared Astronomy Inter-University Cooperation Program, from the Ministry of Education, Culture, Sports, Science and Technology (MEXT) of Japan and the National Research Foundation (NRF) of South Africa. 
We are grateful for observational assistance provided by Noriyuki Matsunaga, Yoshifusa Ita, Enrico Olivier and Shogo Nishiyama. We also thank Martha Boyer and colleagues for allowing us to quote their HST result for V1 in advance of publication and Martha for helpful discussion.

\begin{table*}
\centering
\caption{Long period Miras in LMC, with $(J-K_S)< 1.7$ and that are $>0.8$ mag fainter than the \citet{Ita2011} $K_S$ PLR.}
\begin{tabular}{lllccccccc}
LMC &OGLE&2MASS & $J$ &  $H$ &  $K$ &  $[3.6]$ &  $[4.5]$  & P & $M_K$\\
1 & 5878 &04534486-6857593 & 10.401 & 9.477 & 8.864 & 7.50 &  7.28  & 912 & --9.64\\
2 & 38175 &05175887-6939231 & 10.692 & 10.073 & 9.105  & 8.17 & 7.98 & 830 &--9.39\\
3 & 11637 & 05010330-6854257 & 11.048  &10.073 & 9.355  & 7.69 & 7.60 & 928&  --9.14\\
\end{tabular}\label{LMC}
\end{table*}

\bibliographystyle{mnras}
\bibliography{pawbib.bib}

\begin{thebibliography}{}
\makeatletter
\relax
\def\mn@urlcharsother{\let\do\@makeother \do\$\do\&\do\#\do\^\do\_\do\%\do\~}
\def\mn@doi{\begingroup\mn@urlcharsother \@ifnextchar [ {\mn@doi@}
  {\mn@doi@[]}}
\def\mn@doi@[#1]#2{\def\@tempa{#1}\ifx\@tempa\@empty \href
  {http://dx.doi.org/#2} {doi:#2}\else \href {http://dx.doi.org/#2} {#1}\fi
  \endgroup}
\def\mn@eprint#1#2{\mn@eprint@#1:#2::\@nil}
\def\mn@eprint@arXiv#1{\href {http://arxiv.org/abs/#1} {{\tt arXiv:#1}}}
\def\mn@eprint@dblp#1{\href {http://dblp.uni-trier.de/rec/bibtex/#1.xml}
  {dblp:#1}}
\def\mn@eprint@#1:#2:#3:#4\@nil{\def\@tempa {#1}\def\@tempb {#2}\def\@tempc
  {#3}\ifx \@tempc \@empty \let \@tempc \@tempb \let \@tempb \@tempa \fi \ifx
  \@tempb \@empty \def\@tempb {arXiv}\fi \@ifundefined
  {mn@eprint@\@tempb}{\@tempb:\@tempc}{\expandafter \expandafter \csname
  mn@eprint@\@tempb\endcsname \expandafter{\@tempc}}}

\bibitem[\protect\citeauthoryear{{Anderson}, {Saio}, {Ekstr{\"o}m}, {Georgy}
  \& {Meynet}}{{Anderson} et~al.}{2016}]{Anderson2016}
{Anderson} R.~I.,  {Saio} H.,  {Ekstr{\"o}m} S.,  {Georgy} C.,   {Meynet} G.,
  2016, \mn@doi [\aap] {10.1051/0004-6361/201528031}, \href
  {http://adsabs.harvard.edu/abs/2016A%26A...591A...8A} {591, A8}

\bibitem[\protect\citeauthoryear{{Beccari} et~al.,}{{Beccari}
  et~al.}{2014}]{Beccari2014}
{Beccari} G.,  et~al., 2014, \mn@doi [\aap] {10.1051/0004-6361/201424411},
  \href {http://adsabs.harvard.edu/abs/2014A%26A...570A..78B} {570, A78}

\bibitem[\protect\citeauthoryear{{Bladh}, {H{\"o}fner}, {Aringer}  \&
  {Eriksson}}{{Bladh} et~al.}{2015}]{Bladh2015}
{Bladh} S.,  {H{\"o}fner} S.,  {Aringer} B.,   {Eriksson} K.,  2015, \mn@doi
  [\aap] {10.1051/0004-6361/201424917}, \href
  {http://adsabs.harvard.edu/abs/2015A%26A...575A.105B} {575, A105}

\bibitem[\protect\citeauthoryear{{Bloecker}}{{Bloecker}}{1995}]{Bloecker1995}
{Bloecker} T.,  1995, \aap, \href
  {http://adsabs.harvard.edu/abs/1995A%26A...297..727B} {297, 727}

\bibitem[\protect\citeauthoryear{{Boroson} \& {Liebert}}{{Boroson} \&
  {Liebert}}{1989}]{Boroson1989}
{Boroson} T.~A.,  {Liebert} J.,  1989, \mn@doi [\apj] {10.1086/167340}, \href
  {http://adsabs.harvard.edu/abs/1989ApJ...339..844B} {339, 844}

\bibitem[\protect\citeauthoryear{{Boyer} et~al.,}{{Boyer}
  et~al.}{2015a}]{Boyer2015a}
{Boyer} M.~L.,  et~al., 2015a, \mn@doi [\apjs] {10.1088/0067-0049/216/1/10},
  \href {http://adsabs.harvard.edu/abs/2015ApJS..216...10B} {216, 10}

\bibitem[\protect\citeauthoryear{{Boyer} et~al.,}{{Boyer}
  et~al.}{2015b}]{Boyer2015b}
{Boyer} M.~L.,  et~al., 2015b, \mn@doi [\apj] {10.1088/0004-637X/800/1/51},
  \href {http://adsabs.harvard.edu/abs/2015ApJ...800...51B} {800, 51}

\bibitem[\protect\citeauthoryear{{Bressan}, {Granato}  \& {Silva}}{{Bressan}
  et~al.}{1998}]{Bressan1998}
{Bressan} A.,  {Granato} G.~L.,   {Silva} L.,  1998, \aap, \href
  {http://adsabs.harvard.edu/abs/1998A%26A...332..135B} {332, 135}

\bibitem[\protect\citeauthoryear{{Bressan}, {Marigo}, {Girardi}, {Salasnich},
  {Dal Cero}, {Rubele}  \& {Nanni}}{{Bressan} et~al.}{2012}]{Bressan2012}
{Bressan} A.,  {Marigo} P.,  {Girardi} L.,  {Salasnich} B.,  {Dal Cero} C.,
  {Rubele} S.,   {Nanni} A.,  2012, \mn@doi [\mnras]
  {10.1111/j.1365-2966.2012.21948.x}, \href
  {http://adsabs.harvard.edu/abs/2012MNRAS.427..127B} {427, 127}

\bibitem[\protect\citeauthoryear{{Caffau}, {Ludwig}, {Steffen}, {Freytag}  \&
  {Bonifacio}}{{Caffau} et~al.}{2011}]{Caffau2011}
{Caffau} E.,  {Ludwig} H.-G.,  {Steffen} M.,  {Freytag} B.,   {Bonifacio} P.,
  2011, \mn@doi [\solphys] {10.1007/s11207-010-9541-4}, \href
  {http://adsabs.harvard.edu/abs/2011SoPh..268..255C} {268, 255}

\bibitem[\protect\citeauthoryear{{Carpenter}}{{Carpenter}}{2001}]{Carpenter2001}
{Carpenter} J.~M.,  2001, \mn@doi [\aj] {10.1086/320383}, \href
  {http://adsabs.harvard.edu/abs/2001AJ....121.2851C} {121, 2851}

\bibitem[\protect\citeauthoryear{{Cioni}, {Marquette}, {Loup}, {Azzopardi},
  {Habing}, {Lasserre}  \& {Lesquoy}}{{Cioni} et~al.}{2001}]{Cioni2001}
{Cioni} M.-R.~L.,  {Marquette} J.-B.,  {Loup} C.,  {Azzopardi} M.,  {Habing}
  H.~J.,  {Lasserre} T.,   {Lesquoy} E.,  2001, \mn@doi [\aap]
  {10.1051/0004-6361:20011143}, \href
  {http://adsabs.harvard.edu/abs/2001A%26A...377..945C} {377, 945}

\bibitem[\protect\citeauthoryear{{Cook} \& {Aaronson}}{{Cook} \&
  {Aaronson}}{1988}]{Cook1988}
{Cook} K.~H.,  {Aaronson} M.,  1988, \mn@doi [\pasp] {10.1086/132387}, \href
  {http://adsabs.harvard.edu/abs/1988PASP..100R1218C} {100, 1218}

\bibitem[\protect\citeauthoryear{{Cranmer} \& {Saar}}{{Cranmer} \&
  {Saar}}{2011}]{Cranmer2011}
{Cranmer} S.~R.,  {Saar} S.~H.,  2011, \mn@doi [\apj]
  {10.1088/0004-637X/741/1/54}, \href
  {http://adsabs.harvard.edu/abs/2011ApJ...741...54C} {741, 54}

\bibitem[\protect\citeauthoryear{{Cutri} \& {et al.}}{{Cutri} \& {et
  al.}}{2012}]{Cutri2012}
{Cutri} R.~M.,  {et al.} 2012, VizieR Online Data Catalog, \href
  {http://adsabs.harvard.edu/abs/2012yCat.2307....0C} {2307}

\bibitem[\protect\citeauthoryear{{Cutri} et~al.,}{{Cutri}
  et~al.}{2003}]{Cutri2003}
{Cutri} R.~M.,  et~al., 2003, VizieR Online Data Catalog, \href
  {http://adsabs.harvard.edu/abs/2003yCat.2246....0C} {2246}

\bibitem[\protect\citeauthoryear{{Demers} \& {Battinelli}}{{Demers} \&
  {Battinelli}}{2002}]{Demers2002}
{Demers} S.,  {Battinelli} P.,  2002, \mn@doi [\aj] {10.1086/324735}, \href
  {http://adsabs.harvard.edu/abs/2002AJ....123..238D} {123, 238}

\bibitem[\protect\citeauthoryear{{Doherty}, {Gil-Pons}, {Siess}  \&
  {Lattanzio}}{{Doherty} et~al.}{2017}]{Doherty2017}
{Doherty} C.~L.,  {Gil-Pons} P.,  {Siess} L.,   {Lattanzio} J.~C.,  2017,
  preprint, \href {http://adsabs.harvard.edu/abs/2017arXiv170306895D} {}
  (\mn@eprint {arXiv} {1703.06895})

\bibitem[\protect\citeauthoryear{{Feast}}{{Feast}}{2013}]{Feast2013}
{Feast} M.~W.,  2013, {Galactic Distance Scales}.
p.~829, \mn@doi{10.1007/978-94-007-5612-0_16}

\bibitem[\protect\citeauthoryear{{Feast} \& {Whitelock}}{{Feast} \&
  {Whitelock}}{2000}]{Feast2000}
{Feast} M.~W.,  {Whitelock} P.~A.,  2000, \mn@doi [\mnras]
  {10.1046/j.1365-8711.2000.03629.x}, \href
  {http://adsabs.harvard.edu/abs/2000MNRAS.317..460F} {317, 460}

\bibitem[\protect\citeauthoryear{{Feast} \& {Whitelock}}{{Feast} \&
  {Whitelock}}{2014}]{Feast2014}
{Feast} M.,  {Whitelock} P.~A.,  2014, in {Feltzing} S.,  {Zhao} G.,  {Walton}
  N.~A.,   {Whitelock} P.,  eds,  IAU Symposium Vol. 298, Setting the scene for
  Gaia and LAMOST. pp 40--52 (\mn@eprint {arXiv} {1310.3928}),
  \mn@doi{10.1017/S1743921313006182}

\bibitem[\protect\citeauthoryear{{Feast}, {Glass}, {Whitelock}  \&
  {Catchpole}}{{Feast} et~al.}{1989}]{Feast1989}
{Feast} M.~W.,  {Glass} I.~S.,  {Whitelock} P.~A.,   {Catchpole} R.~M.,  1989,
  \mn@doi [\mnras] {10.1093/mnras/241.3.375}, \href
  {http://adsabs.harvard.edu/abs/1989MNRAS.241..375F} {241, 375}

\bibitem[\protect\citeauthoryear{{Feast}, {Whitelock}  \& {Menzies}}{{Feast}
  et~al.}{2002}]{Feast2002}
{Feast} M.,  {Whitelock} P.,   {Menzies} J.,  2002, \mn@doi [\mnras]
  {10.1046/j.1365-8711.2002.05126.x}, \href
  {http://adsabs.harvard.edu/abs/2002MNRAS.329L...7F} {329, L7}

\bibitem[\protect\citeauthoryear{{Feast}, {Whitelock}  \& {Menzies}}{{Feast}
  et~al.}{2006}]{Feast2006}
{Feast} M.~W.,  {Whitelock} P.~A.,   {Menzies} J.~W.,  2006, \mn@doi [\mnras]
  {10.1111/j.1365-2966.2006.10324.x}, \href
  {http://adsabs.harvard.edu/abs/2006MNRAS.369..791F} {369, 791}

\bibitem[\protect\citeauthoryear{{Girardi}, {Groenewegen}, {Hatziminaoglou}  \&
  {da Costa}}{{Girardi} et~al.}{2005}]{Girardi2005}
{Girardi} L.,  {Groenewegen} M.~A.~T.,  {Hatziminaoglou} E.,   {da Costa} L.,
  2005, \mn@doi [\aap] {10.1051/0004-6361:20042352}, \href
  {http://adsabs.harvard.edu/abs/2005A%26A...436..895G} {436, 895}

\bibitem[\protect\citeauthoryear{{Glass} \& {Lloyd Evans}}{{Glass} \& {Lloyd
  Evans}}{2003}]{Glass2003}
{Glass} I.~S.,  {Lloyd Evans} T.,  2003, \mn@doi [\mnras]
  {10.1046/j.1365-8711.2003.06632.x}, \href
  {http://adsabs.harvard.edu/abs/2003MNRAS.343...67G} {343, 67}

\bibitem[\protect\citeauthoryear{{Gullieuszik}, {Rejkuba}, {Cioni}, {Habing}
  \& {Held}}{{Gullieuszik} et~al.}{2007}]{Gullieuszik2007}
{Gullieuszik} M.,  {Rejkuba} M.,  {Cioni} M.~R.,  {Habing} H.~J.,   {Held}
  E.~V.,  2007, \mn@doi [\aap] {10.1051/0004-6361:20066848}, \href
  {http://adsabs.harvard.edu/abs/2007A%26A...475..467G} {475, 467}

\bibitem[\protect\citeauthoryear{{Higgs} et~al.,}{{Higgs}
  et~al.}{2016}]{Higgs2016}
{Higgs} C.~R.,  et~al., 2016, \mn@doi [\mnras] {10.1093/mnras/stw257}, \href
  {http://adsabs.harvard.edu/abs/2016MNRAS.458.1678H} {458, 1678}

\bibitem[\protect\citeauthoryear{{H{\"o}fner}, {Bladh}, {Aringer}  \&
  {Ahuja}}{{H{\"o}fner} et~al.}{2016}]{Hofner2016}
{H{\"o}fner} S.,  {Bladh} S.,  {Aringer} B.,   {Ahuja} R.,  2016, \mn@doi
  [\aap] {10.1051/0004-6361/201628424}, \href
  {http://adsabs.harvard.edu/abs/2016A%26A...594A.108H} {594, A108}

\bibitem[\protect\citeauthoryear{{Hughes} \& {Wood}}{{Hughes} \&
  {Wood}}{1990}]{Hughes1990}
{Hughes} S.~M.~G.,  {Wood} P.~R.,  1990, \mn@doi [\aj] {10.1086/115374}, \href
  {http://adsabs.harvard.edu/abs/1990AJ.....99..784H} {99, 784}

\bibitem[\protect\citeauthoryear{{Ita} \& {Matsunaga}}{{Ita} \&
  {Matsunaga}}{2011}]{Ita2011}
{Ita} Y.,  {Matsunaga} N.,  2011, \mn@doi [\mnras]
  {10.1111/j.1365-2966.2010.18056.x}, \href
  {http://adsabs.harvard.edu/abs/2011MNRAS.412.2345I} {412, 2345}

\bibitem[\protect\citeauthoryear{{Ita} et~al.,}{{Ita} et~al.}{2004}]{Ita2004}
{Ita} Y.,  et~al., 2004, \mn@doi [\mnras] {10.1111/j.1365-2966.2004.08126.x},
  \href {http://adsabs.harvard.edu/abs/2004MNRAS.353..705I} {353, 705}

\bibitem[\protect\citeauthoryear{{Jacoby}}{{Jacoby}}{1980}]{Jacoby1980}
{Jacoby} G.~H.,  1980, \mn@doi [\apjs] {10.1086/190642}, \href
  {http://adsabs.harvard.edu/abs/1980ApJS...42....1J} {42, 1}

\bibitem[\protect\citeauthoryear{{Karakas}, {Lattanzio}  \& {Pols}}{{Karakas}
  et~al.}{2002}]{Karakas2002}
{Karakas} A.~I.,  {Lattanzio} J.~C.,   {Pols} O.~R.,  2002, \mn@doi [\pasa]
  {10.1071/AS02013}, \href {http://adsabs.harvard.edu/abs/2002PASA...19..515K}
  {19, 515}

\bibitem[\protect\citeauthoryear{{Karakas}, {Ventura}, {Dell'Agli}  \& {Di
  Criscienzo}}{{Karakas} et~al.}{2017}]{Karakas2017}
{Karakas} A.,  {Ventura} P.,  {Dell'Agli} F.,   {Di Criscienzo} M.,  2017,
  \memsai, \href {http://adsabs.harvard.edu/abs/2016MmSAI..87..225V} {in press}

\bibitem[\protect\citeauthoryear{{Kasliwal} et~al.,}{{Kasliwal}
  et~al.}{2017}]{Kasliwal2017}
{Kasliwal} M.~M.,  et~al., 2017, \mn@doi [\apj] {10.3847/1538-4357/aa6978},
  \href {http://adsabs.harvard.edu/abs/2017ApJ...839...88K} {839, 88}

\bibitem[\protect\citeauthoryear{{Kirby}, {Rizzi}, {Held}, {Cohen}, {Cole},
  {Manning}, {Skillman}  \& {Weisz}}{{Kirby} et~al.}{2017}]{Kirby2017}
{Kirby} E.~N.,  {Rizzi} L.,  {Held} E.~V.,  {Cohen} J.~G.,  {Cole} A.~A.,
  {Manning} E.~M.,  {Skillman} E.~D.,   {Weisz} D.~R.,  2017, \mn@doi [\apj]
  {10.3847/1538-4357/834/1/9}, \href
  {http://adsabs.harvard.edu/abs/2017ApJ...834....9K} {834, 9}

\bibitem[\protect\citeauthoryear{{Macri}, {Ngeow}, {Kanbur}, {Mahzooni}  \&
  {Smitka}}{{Macri} et~al.}{2015}]{Macri2015}
{Macri} L.~M.,  {Ngeow} C.-C.,  {Kanbur} S.~M.,  {Mahzooni} S.,   {Smitka}
  M.~T.,  2015, \mn@doi [\aj] {10.1088/0004-6256/149/4/117}, \href
  {http://adsabs.harvard.edu/abs/2015AJ....149..117M} {149, 117}

\bibitem[\protect\citeauthoryear{{Marigo}, {Girardi}, {Bressan}, {Groenewegen},
  {Silva}  \& {Granato}}{{Marigo} et~al.}{2008}]{Marigo2008}
{Marigo} P.,  {Girardi} L.,  {Bressan} A.,  {Groenewegen} M.~A.~T.,  {Silva}
  L.,   {Granato} G.~L.,  2008, \mn@doi [\aap] {10.1051/0004-6361:20078467},
  \href {http://adsabs.harvard.edu/abs/2008A%26A...482..883M} {482, 883}

\bibitem[\protect\citeauthoryear{{Marigo}, {Bressan}, {Nanni}, {Girardi}  \&
  {Pumo}}{{Marigo} et~al.}{2013}]{Marigo2013}
{Marigo} P.,  {Bressan} A.,  {Nanni} A.,  {Girardi} L.,   {Pumo} M.~L.,  2013,
  \mn@doi [\mnras] {10.1093/mnras/stt1034}, \href
  {http://adsabs.harvard.edu/abs/2013MNRAS.434..488M} {434, 488}

\bibitem[\protect\citeauthoryear{{Marigo} et~al.,}{{Marigo}
  et~al.}{2017}]{Marigo2017}
{Marigo} P.,  et~al., 2017, \mn@doi [\apj] {10.3847/1538-4357/835/1/77}, \href
  {http://adsabs.harvard.edu/abs/2017ApJ...835...77M} {835, 77}

\bibitem[\protect\citeauthoryear{{Matsunaga} et~al.,}{{Matsunaga}
  et~al.}{2013}]{Matsunaga2013}
{Matsunaga} N.,  et~al., 2013, \mn@doi [\mnras] {10.1093/mnras/sts343}, \href
  {http://adsabs.harvard.edu/abs/2013MNRAS.429..385M} {429, 385}

\bibitem[\protect\citeauthoryear{{Mattsson}, {Wahlin}  \&
  {H{\"o}fner}}{{Mattsson} et~al.}{2010}]{Mattsson2010}
{Mattsson} L.,  {Wahlin} R.,   {H{\"o}fner} S.,  2010, \mn@doi [\aap]
  {10.1051/0004-6361/200912084}, \href
  {http://adsabs.harvard.edu/abs/2010A%26A...509A..14M} {509, A14}

\bibitem[\protect\citeauthoryear{{McConnachie}}{{McConnachie}}{2012}]{McConnachie2012}
{McConnachie} A.~W.,  2012, \mn@doi [\aj] {10.1088/0004-6256/144/1/4}, \href
  {http://adsabs.harvard.edu/abs/2012AJ....144....4M} {144, 4}

\bibitem[\protect\citeauthoryear{{McQuinn} et~al.,}{{McQuinn}
  et~al.}{2017}]{McQuinn2017}
{McQuinn} K.~B.~W.,  et~al., 2017, \mn@doi [\apj] {10.3847/1538-4357/834/1/78},
  \href {http://adsabs.harvard.edu/abs/2017ApJ...834...78M} {834, 78}

\bibitem[\protect\citeauthoryear{{Menzies}, {Feast}, {Tanab{\'e}}, {Whitelock}
  \& {Nakada}}{{Menzies} et~al.}{2002}]{Menzies2002}
{Menzies} J.,  {Feast} M.,  {Tanab{\'e}} T.,  {Whitelock} P.,   {Nakada} Y.,
  2002, \mn@doi [\mnras] {10.1046/j.1365-8711.2002.05759.x}, \href
  {http://adsabs.harvard.edu/abs/2002MNRAS.335..923M} {335, 923}

\bibitem[\protect\citeauthoryear{{Menzies}, {Feast}, {Whitelock}, {Olivier},
  {Matsunaga}  \& {da Costa}}{{Menzies} et~al.}{2008}]{Menzies2008}
{Menzies} J.,  {Feast} M.,  {Whitelock} P.,  {Olivier} E.,  {Matsunaga} N.,
  {da Costa} G.,  2008, \mn@doi [\mnras] {10.1111/j.1365-2966.2008.12907.x},
  \href {http://adsabs.harvard.edu/abs/2008MNRAS.385.1045M} {385, 1045}

\bibitem[\protect\citeauthoryear{{Menzies}, {Whitelock}, {Feast}  \&
  {Matsunaga}}{{Menzies} et~al.}{2010}]{Menzies2010}
{Menzies} J.~W.,  {Whitelock} P.~A.,  {Feast} M.~W.,   {Matsunaga} N.,  2010,
  \mn@doi [\mnras] {10.1111/j.1365-2966.2010.16670.x}, \href
  {http://adsabs.harvard.edu/abs/2010MNRAS.406...86M} {406, 86}

\bibitem[\protect\citeauthoryear{{Menzies}, {Feast}, {Whitelock}  \&
  {Matsunaga}}{{Menzies} et~al.}{2011}]{Menzies2011}
{Menzies} J.~W.,  {Feast} M.~W.,  {Whitelock} P.~A.,   {Matsunaga} N.,  2011,
  \mn@doi [\mnras] {10.1111/j.1365-2966.2011.18649.x}, \href
  {http://adsabs.harvard.edu/abs/2011MNRAS.414.3492M} {414, 3492}

\bibitem[\protect\citeauthoryear{{Menzies}, {Whitelock}  \& {Feast}}{{Menzies}
  et~al.}{2015}]{Menzies2015}
{Menzies} J.~W.,  {Whitelock} P.~A.,   {Feast} M.~W.,  2015, \mn@doi [\mnras]
  {10.1093/mnras/stv1310}, \href
  {http://adsabs.harvard.edu/abs/2015MNRAS.452..910M} {452, 910}

\bibitem[\protect\citeauthoryear{{Mohamed} \& {Podsiadlowski}}{{Mohamed} \&
  {Podsiadlowski}}{2012}]{Mohamed2012}
{Mohamed} S.,  {Podsiadlowski} P.,  2012, Baltic Astronomy, \href
  {http://adsabs.harvard.edu/abs/2012BaltA..21...88M} {21, 88}

\bibitem[\protect\citeauthoryear{{Momany} et~al.,}{{Momany}
  et~al.}{2005}]{Momany2005}
{Momany} Y.,  et~al., 2005, \mn@doi [\aap] {10.1051/0004-6361:20052747}, \href
  {http://adsabs.harvard.edu/abs/2005A%26A...439..111M} {439, 111}

\bibitem[\protect\citeauthoryear{{Momany} et~al.,}{{Momany}
  et~al.}{2014}]{Momany2014}
{Momany} Y.,  et~al., 2014, \mn@doi [\aap] {10.1051/0004-6361/201424055}, \href
  {http://adsabs.harvard.edu/abs/2014A%26A...572A..42M} {572, A42}

\bibitem[\protect\citeauthoryear{{Nanni}, {Marigo}, {Groenewegen}, {Aringer},
  {Girardi}, {Pastorelli}, {Bressan}  \& {Bladh}}{{Nanni}
  et~al.}{2016}]{Nanni2016}
{Nanni} A.,  {Marigo} P.,  {Groenewegen} M.~A.~T.,  {Aringer} B.,  {Girardi}
  L.,  {Pastorelli} G.,  {Bressan} A.,   {Bladh} S.,  2016, \mn@doi [\mnras]
  {10.1093/mnras/stw1681}, \href
  {http://adsabs.harvard.edu/abs/2016MNRAS.462.1215N} {462, 1215}

\bibitem[\protect\citeauthoryear{{Nikutta}, {Hunt-Walker}, {Nenkova},
  {Ivezi{\'c}}  \& {Elitzur}}{{Nikutta} et~al.}{2014}]{Nikutta2014}
{Nikutta} R.,  {Hunt-Walker} N.,  {Nenkova} M.,  {Ivezi{\'c}} {\v Z}.,
  {Elitzur} M.,  2014, \mn@doi [\mnras] {10.1093/mnras/stu1087}, \href
  {http://adsabs.harvard.edu/abs/2014MNRAS.442.3361N} {442, 3361}

\bibitem[\protect\citeauthoryear{{Riebel} et~al.,}{{Riebel}
  et~al.}{2015}]{Riebel2015}
{Riebel} D.,  et~al., 2015, \mn@doi [\apj] {10.1088/0004-637X/807/1/1}, \href
  {http://adsabs.harvard.edu/abs/2015ApJ...807....1R} {807, 1}

\bibitem[\protect\citeauthoryear{{Saviane}, {Rizzi}, {Held}, {Bresolin}  \&
  {Momany}}{{Saviane} et~al.}{2002}]{Saviane2002}
{Saviane} I.,  {Rizzi} L.,  {Held} E.~V.,  {Bresolin} F.,   {Momany} Y.,  2002,
  \mn@doi [\aap] {10.1051/0004-6361:20020750}, \href
  {http://adsabs.harvard.edu/abs/2002A%26A...390...59S} {390, 59}

\bibitem[\protect\citeauthoryear{{Schlafly} et~al.,}{{Schlafly}
  et~al.}{2016}]{Schlafly2016}
{Schlafly} E.~F.,  et~al., 2016, \mn@doi [\apj] {10.3847/0004-637X/821/2/78},
  \href {http://adsabs.harvard.edu/abs/2016ApJ...821...78S} {821, 78}

\bibitem[\protect\citeauthoryear{{Sloan} et~al.,}{{Sloan}
  et~al.}{2012}]{Sloan2012}
{Sloan} G.~C.,  et~al., 2012, \mn@doi [\apj] {10.1088/0004-637X/752/2/140},
  \href {http://adsabs.harvard.edu/abs/2012ApJ...752..140S} {752, 140}

\bibitem[\protect\citeauthoryear{{Soszy{\'n}ski} et~al.,}{{Soszy{\'n}ski}
  et~al.}{2009}]{Soszynski2009}
{Soszy{\'n}ski} I.,  et~al., 2009, \actaa, \href
  {http://adsabs.harvard.edu/abs/2009AcA....59..239S} {59, 239}

\bibitem[\protect\citeauthoryear{{Totten}, {Irwin}  \& {Whitelock}}{{Totten}
  et~al.}{2000}]{Totten2000}
{Totten} E.~J.,  {Irwin} M.~J.,   {Whitelock} P.~A.,  2000, \mn@doi [\mnras]
  {10.1046/j.1365-8711.2000.03370.x}, \href
  {http://adsabs.harvard.edu/abs/2000MNRAS.314..630T} {314, 630}

\bibitem[\protect\citeauthoryear{{Weisz}, {Dolphin}, {Skillman}, {Holtzman},
  {Gilbert}, {Dalcanton}  \& {Williams}}{{Weisz} et~al.}{2014}]{Weisz2014}
{Weisz} D.~R.,  {Dolphin} A.~E.,  {Skillman} E.~D.,  {Holtzman} J.,  {Gilbert}
  K.~M.,  {Dalcanton} J.~J.,   {Williams} B.~F.,  2014, \mn@doi [\apj]
  {10.1088/0004-637X/789/2/147}, \href
  {http://adsabs.harvard.edu/abs/2014ApJ...789..147W} {789, 147}

\bibitem[\protect\citeauthoryear{{Whitelock}}{{Whitelock}}{1987}]{Whitelock1987}
{Whitelock} P.~A.,  1987, \mn@doi [\pasp] {10.1086/132019}, \href
  {http://adsabs.harvard.edu/abs/1987PASP...99..573W} {99, 573}

\bibitem[\protect\citeauthoryear{{Whitelock}}{{Whitelock}}{2013}]{Whitelock2013b}
{Whitelock} P.~A.,  2013, in {de Grijs} R.,  ed.,  IAU Symposium Vol. 289,
  Advancing the Physics of Cosmic Distances. pp 209--216 (\mn@eprint {arXiv}
  {1210.7307}), \mn@doi{10.1017/S1743921312021400}

\bibitem[\protect\citeauthoryear{{Whitelock} \& {Feast}}{{Whitelock} \&
  {Feast}}{2014}]{Whitelock2014}
{Whitelock} P.~A.,  {Feast} M.~W.,  2014, in EAS Publications Series. pp
  263--269, \mn@doi{10.1051/eas/1567047}

\bibitem[\protect\citeauthoryear{{Whitelock}, {Feast}, {van Loon}  \&
  {Zijlstra}}{{Whitelock} et~al.}{2003}]{Whitelock2003}
{Whitelock} P.~A.,  {Feast} M.~W.,  {van Loon} J.~T.,   {Zijlstra} A.~A.,
  2003, \mn@doi [\mnras] {10.1046/j.1365-8711.2003.06514.x}, \href
  {http://adsabs.harvard.edu/abs/2003MNRAS.342...86W} {342, 86}

\bibitem[\protect\citeauthoryear{{Whitelock}, {Feast}, {Marang}  \&
  {Groenewegen}}{{Whitelock} et~al.}{2006}]{Whitelock2006}
{Whitelock} P.~A.,  {Feast} M.~W.,  {Marang} F.,   {Groenewegen} M.~A.~T.,
  2006, \mn@doi [\mnras] {10.1111/j.1365-2966.2006.10322.x}, \href
  {http://adsabs.harvard.edu/abs/2006MNRAS.369..751W} {369, 751}

\bibitem[\protect\citeauthoryear{{Whitelock}, {Feast}  \& {van
  Leeuwen}}{{Whitelock} et~al.}{2008}]{Whitelock2008}
{Whitelock} P.~A.,  {Feast} M.~W.,   {van Leeuwen} F.,  2008, \mn@doi [\mnras]
  {10.1111/j.1365-2966.2008.13032.x}, \href
  {http://adsabs.harvard.edu/abs/2008MNRAS.386..313W} {386, 313}

\bibitem[\protect\citeauthoryear{{Whitelock}, {Menzies}, {Feast}, {Matsunaga},
  {Tanab{\'e}}  \& {Ita}}{{Whitelock} et~al.}{2009}]{Whitelock2009}
{Whitelock} P.~A.,  {Menzies} J.~W.,  {Feast} M.~W.,  {Matsunaga} N.,
  {Tanab{\'e}} T.,   {Ita} Y.,  2009, \mn@doi [\mnras]
  {10.1111/j.1365-2966.2008.14365.x}, \href
  {http://adsabs.harvard.edu/abs/2009MNRAS.394..795W} {394, 795}

\bibitem[\protect\citeauthoryear{{Whitelock}, {Menzies}, {Feast}, {Nsengiyumva}
   \& {Matsunaga}}{{Whitelock} et~al.}{2013}]{Whitelock2013}
{Whitelock} P.~A.,  {Menzies} J.~W.,  {Feast} M.~W.,  {Nsengiyumva} F.,
  {Matsunaga} N.,  2013, \mn@doi [\mnras] {10.1093/mnras/sts188}, \href
  {http://adsabs.harvard.edu/abs/2013MNRAS.428.2216W} {428, 2216}

\bibitem[\protect\citeauthoryear{{Whitelock}, {Kasliwal}  \&
  {Boyer}}{{Whitelock} et~al.}{2017}]{Whitelock2017}
{Whitelock} P.~A.,  {Kasliwal} M.,   {Boyer} M.,  2017, to appear in: (eds.)
  Catalan M., Gieren W., "Wide-Field Variability Surveys: A 21st Century
  Perspective"; arXiv:1702.06797, \href
  {http://adsabs.harvard.edu/abs/2017arXiv170206797W} {}

\bibitem[\protect\citeauthoryear{{Wood}}{{Wood}}{2015}]{Wood2015}
{Wood} P.~R.,  2015, \mn@doi [\mnras] {10.1093/mnras/stv289}, \href
  {http://adsabs.harvard.edu/abs/2015MNRAS.448.3829W} {448, 3829}

\bibitem[\protect\citeauthoryear{{Wood} et~al.,}{{Wood}
  et~al.}{1999}]{Wood1999}
{Wood} P.~R.,  et~al., 1999, in {Le Bertre} T.,  {Lebre} A.,   {Waelkens} C.,
  eds,  IAU Symposium Vol. 191, Asymptotic Giant Branch Stars. p.~151

\bibitem[\protect\citeauthoryear{{Yuan}, {He}, {Macri}, {Long}  \&
  {Huang}}{{Yuan} et~al.}{2017}]{Yuan2017}
{Yuan} W.,  {He} S.,  {Macri} L.~M.,  {Long} J.,   {Huang} J.~Z.,  2017,
  \mn@doi [\aj] {10.3847/1538-3881/aa63f1}, \href
  {http://adsabs.harvard.edu/abs/2017AJ....153..170Y} {153, 170}

\makeatother
\end{thebibliography}

\section{Appendix: LMC Sources that are potentially similar to V1}\label{appendix}
It is important to establish if there are sources elsewhere with similar characteristics to V1 in Sgr~dIG. We therefore examined O-rich Miras in the LMC with periods over 600 days, as identified by OGLE \citep{Soszynski2009}, together with $JHK_S$ photometry from 2MASS \citep{Cutri2003}. There are only nine stars with $K_S$ magnitudes that are 0.8 mag or more below the \citet{Ita2011} PLR. Six of these are red ($(J-K_S)>1.9$) and therefore probably affected by circumstellar extinction. The other 3 are listed in Table~\ref{LMC} and are referred to below as LMC1, 2 and 3; they all fall close to V1 in Fig.~\ref{plk}.  All three have good OGLE light curves showing the bumps on the rising branch that are common in luminous long period Miras \citep{Glass2003}. 

LMC2 is near the centre of the LMC and has been better studied than the other two sources. Although it was originally classified as a planetary nebula, Jacoby LMC 19 \citep{Jacoby1980}, this was not confirmed \citep{Boroson1989} and it is clearly an AGB variable. It has extra published $JHK_S$ photometry \citep{Macri2015} that ranges in $K_S$ from 7.79 to 8.46 mag. This shows that the 2MASS measure was particularly faint, the $K_S$ amplitude is $>1.3$ mag and the mean magnitude lies very close to the PLR. Multi-epoch IRSF photometry from Ita (private communication) confirms the period and indicates that $K_S \sim 9.1$ corresponds to minimum light. LMC3 has $K_S=8.39$ and 8.65 in the DENIS catalogue showing that the 2MASS value is probably nearer the minimum than the mean. The DENIS values for LMC1, $K_S=8.98$ and 8.86, are very similar to those from 2MASS. 

Thus we are left with only a single star in the LMC that might be like V1 in terms of its near-infrared character. The Spitzer magnitudes for LMC1 differ from the \citet{Yuan2017} and \citet{Ita2011} relations by 0.15 and -0.13 mag, respectively, at [3.5] and 0.07 and -0.26 mag, at [4.5], which suggest it is probably a normal HBB star. More observations are required to confirm this, but there are no strong candidates in the LMC that look like V1 in Sgr~dIG.

\end{document}